%% file: main.tex
\documentclass[comsoc]{IEEEtran}
\usepackage{graphicx}
\usepackage{caption}
\usepackage{subcaption}
\usepackage{epstopdf}
\usepackage{booktabs}
\usepackage{multirow}
\usepackage{amsmath}
\usepackage{amssymb}
\usepackage{amsthm}
\usepackage{url}
\usepackage[numbers,square,comma,compress,sort]{natbib}
\usepackage[ruled,linesnumbered]{algorithm2e}
\usepackage{algorithmicx}
\usepackage{algpseudocode}

\usepackage{todonotes} 

\DeclareMathOperator*{\argmin}{\arg\!\min}

\begin{document}
%
\title{An Online Learning Methodology for Performance Modeling of Graphics Processors}

\author{Ujjwal Gupta, Manoj Babu, Raid Ayoub, Michael Kishinevsky, \\Francesco Paterna, Suat Gumussoy, and Umit Y. Ogras

\IEEEcompsocitemizethanks{
\IEEEcompsocthanksitem U. Gupta, M. Babu and U. Y. Ogras are with the
School of Electrical, Computer, and Energy Engineering, Arizona State University, Tempe, AZ, 85281.
E-mail: \{ujjwal, msbabu, umit\}@asu.edu
\IEEEcompsocthanksitem R. Ayoub, M. Kishinevsky and F. Paterna are with Intel Corporation.
E-mail: \{raid.ayoub, michael.kishinevsky, francesco.paterna\}@intel.com%
\IEEEcompsocthanksitem S. Gumussoy is an IEEE member.
E-mail: suat@gumussoy.net }
\thanks{Manuscript received September 13, 2017.}

}

\maketitle

\input{abstract.tex}

\begin{IEEEkeywords}
Integrated GPUs, performance modeling, online learning, frequency scaling, power management, RLS
\end{IEEEkeywords}

\input{introduction.tex}
\input{related_research.tex}

\input{motivation.tex}
\input{methodology.tex}

\input{experiments.tex}

\input{conclusion_future_research.tex}

\bibliographystyle{IEEEtranSN}
{\footnotesize
\bibliography{main_embedded_refs}}

\begin{IEEEbiography}[{\includegraphics[width=1in,height=1.25in,clip,keepaspectratio]{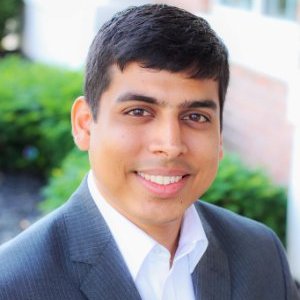}}]%
{Ujjwal Gupta} received the B.E. degree from Manipal University, KA, India, in 2011, in Electronics and Communication Engineering, and the M.S. degree from Arizona State University, Tempe, AZ, in 2014, in Electrical Engineering. He is currently working toward the Ph.D. degree in the Department of Electrical, Computer, and Energy Engineering, Arizona State University, Tempe. His research interests include dynamic power management and electronic design automation of Multiprocessor Systems-on-Chips.
\end{IEEEbiography}

\vspace{-1cm}

\begin{IEEEbiography}[{\includegraphics[width=1in,height=1.25in,clip,keepaspectratio]{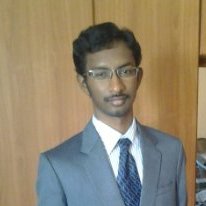}}]%
{Manoj Babu} received his B.E. in Telecommunication Engineering from PESIT, Bangalore, India.
He is now pursuing a Ph.D. in Computer Engineering at ASU. Manoj has previously worked for a
year as a platform engineer at Nvidia in India.
\end{IEEEbiography}

\vspace{-1cm}

\begin{IEEEbiography}[{\includegraphics[width=1in,height=1.25in,clip,keepaspectratio]{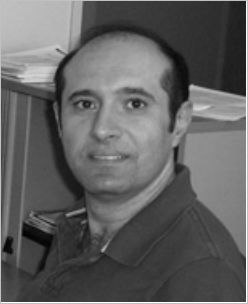}}]%
{Raid Ayoub} (S’09) received the B.S. and M.S. degrees
in electrical engineering from the University of
Technology, Baghdad, Iraq and a Ph.D. degree in computer engineering from the
department of computer science and engineering, University of California at San Diego, La Jolla in 2011.
Currently he is a research scientist at the strategic CAD labs of Intel Corporation.
His current research interests include dynamic control and runtime optimizations at the system level to improve energy efficiency and performance of computing systems.
\end{IEEEbiography}

\vspace{-1cm}

\begin{IEEEbiography}[{\includegraphics[width=1in,height=1.25in,clip,keepaspectratio]{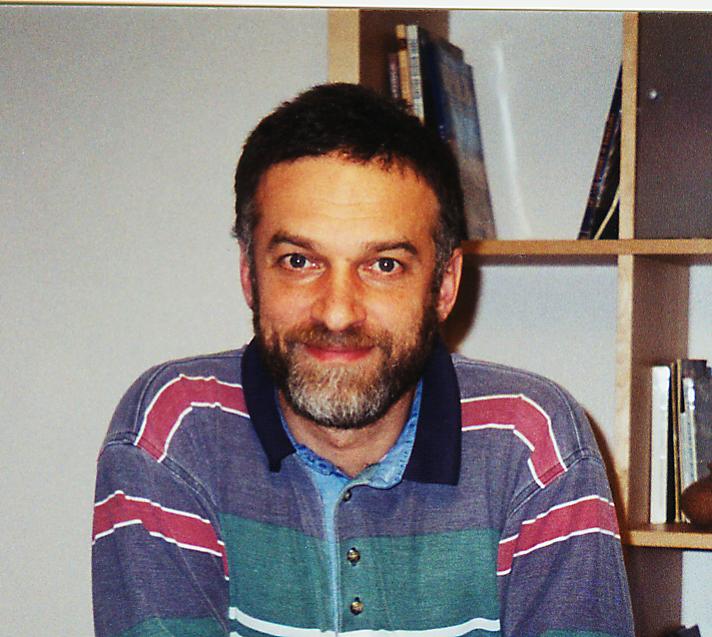}}]%
{Michael Kishinevsky} received the MS and PhD degrees in CS from the Electrotechnical University of St. Petersburg. He leads a research group in system design and architecture at Strategic CAD Labs of Intel. Prior to joining Intel in 1998, he has been a research fellow at the Russian Academy of Science, a senior researcher at a start-up in asynchronous design (TRASSA), a visiting associate professor at the Technical University of Denmark, and a professor at the University of Aizu, Japan. He coauthored three books in asynchronous design and has published more than 100 journal and conference papers. He received the Semiconductor Research Corporation outstanding mentor awards (2004 and 2010) and the best paper awards at the Design Automation Conference (2004) and the International Conference on Application of Concurrency to System Design (2009).
\end{IEEEbiography}

\vspace{-1cm}

\begin{IEEEbiography}[{\includegraphics[width=1in,height=1.25in,clip,keepaspectratio]{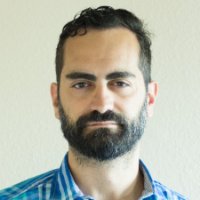}}]%
{Francesco Paterna} received the MS degree
(summa cum laude) in electrical engineering
from the Universita` di Roma La Sapienza, Italy,
in 2007 and the PhD degree in electronics,
computer sciences, and telecommunications
from Universita` di Bologna in 2012. In his PhD
thesis titled “Variability-tolerant High-reliability
Multicore Platforms” he proposed runtime adaptive
techniques to mitigate the impact of process
variations and wear-out in multicore processor
system on chips in terms of performance, energy consumption, and
lifetime. He was a postdoctoral research associate at Brown University
and afterwards at University of California, San Diego. His
works have been published in peer-reviewed international journals and
conferences. Currently, he is working as a Research Scientist at Intel Corporation
within the Strategic CAD Labs.
\end{IEEEbiography}

\vspace{-1cm}

\begin{IEEEbiography}[{\includegraphics[width=1in,height=1.25in,clip,keepaspectratio]{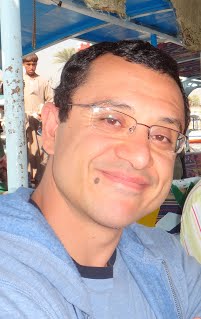}}]%
{Suat Gumussoy} received his B.S. degrees in Electrical \& Electronics Engineering and Mathematics from Middle East Technical University at Turkey in 1999 and M.S., Ph.D. degrees in Electrical and Computer Engineering from The Ohio State University at USA in 2001 and 2004. He worked as a system engineer in electronic self-protection system design for F-16 aircrafts at Mikes, New York (2005-2007). He was a postdoctoral researcher in Computer Science Department at University of Leuven, Belgium (2008-2011). He is currently principal control system engineer in Controls and Identification Team at MathWorks. He develops robust control, system identification and model reduction algorithms. He is the author of two control apps, Control System Tuner for fixed-order, fixed-structure multi loop controller design and Model Reducer for model simplification and reduction. Dr. Gumussoy's research focus is on analysis and control of time delay systems, optimization based tuning methods for robust control, frequency domain system identification and model reduction. He is the author of more than 40 publications and serves as an Associate Editor in IEEE Conference Editorial Board.
\end{IEEEbiography}

\vspace{-1cm}

\begin{IEEEbiography}[{\includegraphics[width=1in,height=1.25in,clip,keepaspectratio]{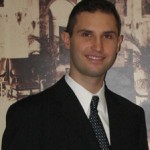}}]%
{Umit Y. Ogras} received his Ph.D. degree in Electrical and Computer Engineering from Carnegie Mellon University, Pittsburgh, PA, in 2007. From 2008 to 2013, he worked as a Research Scientist at the Strategic CAD Laboratories, Intel Corporation. He is currently an Assistant Professor at the School of Electrical, Computer and Energy Engineering. Recognitions Dr. Ogras has received include Strategic CAD Labs Research Award, 2012 IEEE Donald O. Pederson Transactions on CAD Best Paper Award, 2011 IEEE VLSI Transactions Best Paper Award and 2008 EDAA Outstanding PhD. Dissertation Award. His research interests include digital system design, embedded systems, multicore architecture and electronic design automation with particular emphasis on multiprocessor systems-on-chip (MPSoC).
\end{IEEEbiography}

\clearpage
\newpage
\input{appendix1.tex}
\end{document}

%% file: abstract.tex
\begin{abstract}

\textcolor{black}{Approximately 18\% of the 3.2 million smartphone applications rely on integrated graphics processing units (GPUs) to achieve competitive performance. Graphics performance, typically measured in frames per second,
is a strong function of the GPU frequency, which in turn has a significant impact on mobile processor power consumption.}
Consequently, dynamic power management algorithms have to assess the performance sensitivity to the frequency accurately
to choose the operating frequency of the GPU effectively.
Since the impact of GPU frequency on performance varies rapidly over time,
there is a need for online performance models that can adapt to varying workloads.
This paper presents a light-weight adaptive runtime performance model that predicts the frame processing time
of graphics workloads at runtime without apriori characterization.
We employ this model to estimate the frame time sensitivity to the GPU frequency,
\emph{i.e.}, the partial derivative of the frame time with respect to the GPU frequency.
The proposed model does not rely on any parameter learned offline.
Our experiments on the Intel Minnowboard MAX platform running common GPU benchmarks show that the mean absolute percentage error in frame time and frame time sensitivity prediction are 4.2\% and 6.7\%, respectively.
\end{abstract}

%% file: introduction.tex
\section{Introduction} \label{sec:intro}


\textcolor{black}{
Graphically-intensive mobile applications, such as games, constitute about 18\% of the most popular smartphone application categories~\cite{AppTornadoApp}.
Consequently, integrated GPUs have become an indispensable component of mobile processors due to the increasing popularity of graphics applications.
Our measurements show that the GPU power consumption accounts for more than 35$\%$ of application processor power
when running many of these applications.}
The GPU frequency cannot be reduced arbitrarily to save power, since it also determines the achievable frame rate, which has a significant impact on the user experience.
\textcolor{black}{Therefore, there is a growing need to use graphics performance models that can
accurately and judiciously control the GPU frequency.}

\begin{figure}
	\centering
	\includegraphics[width=0.9\linewidth]{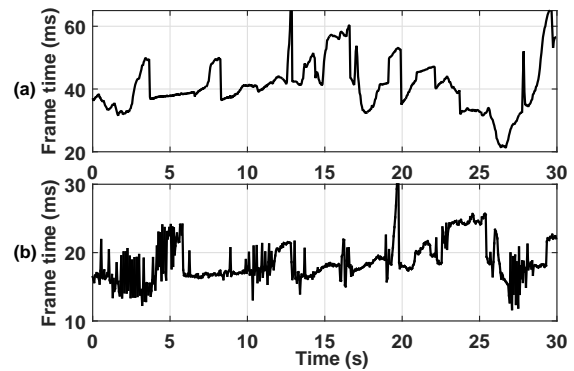}\\
	\caption{The change in frame time for ice-storm application for (a) 200 MHz and (b) 489 MHz GPU frequencies.}\label{fig:intro_motivation}
\end{figure}

The primary graphics performance metric is the number of frames that can be processed per second,
since \textcolor{black}{this} limits the maximum display frame rate.
Therefore, we use the time the GPU takes to process a frame as the performance metric.
\textcolor{black}{Frame time highly correlates with GPU frequency, and is dependent on the target application.}
Furthermore, it varies significantly throughout the lifetime of an application, as shown in Figure~\ref{fig:intro_motivation}.
That is, the frame time is a multivariate function of the frequency and
workload, where the latter is captured by the performance counters.
\textcolor{black}{Therefore, an effective GPU performance model must adapt to the dynamic workload variations to accurately predict the frame time as a function of frequency.}
%

\textcolor{black}{In this paper, we present a performance model that when combined with a power model can be integrated into dynamic
power management algorithms to enable selection of the best GPU frequency for graphics applications.
We develop a systematic two-step methodology for constructing
a tractable runtime model for GPU frame time prediction.}
The first step is an extensive analysis to collect frame time and GPU performance counter data.
This analysis enables us to construct a frame time model template and select the feature set \textcolor{black}{that should be used online}.
Our model employs differential calculus to express the change in frame time
as a function of the partial derivatives of the frame time with respect to the GPU frequency
and performance counters.
\textcolor{black}{In the second step, we implement an adaptive algorithm, whose function is to learn the coefficients of the proposed model online
for dynamically predicting the change in the frame time.}
\textcolor{black}{Unlike our previous work~\cite{Gupta2016Adaptive}, the proposed adaptive algorithm
\textit{does not depend on modeling any performance counters offline}.
%
%
\textcolor{black}{We achieve this by identifying the counters that depend on the GPU frequency during the offline feature selection process.
Hence, we exploit the characterization data, which is already available,
and construct a fully online model without relying on micro-architectural details.}
Furthermore, we present two different online algorithms that can be employed
based on the number of features used in the model.
The first algorithm is the covariance form of recursive least squares (RLS) algorithm.
RLS is a good choice since the correlation between different frames decays quickly
unlike the fractal behavior observed at the macroblock level~\cite{varatkar2004chip}.
The covariance form avoids matrix inversion and
incurs very small overhead when the number of features is small ($\approx \hspace{-1mm}10$).
However, its computational complexity still grows quadratically~\cite{Mendel1995Lessons}.
Therefore, we also employ the traversal form of RLS with coordinate descent, called Dichotomous Coordinate Descent form of RLS (DCD-RLS),
whose complexity grows linearly with the number of features~\cite{Zakharov2008Low}.
We employ the adaptive frame time model to estimate the frame time sensitivity to the GPU frequency,
which is defined as the partial derivative of the frame time with respect to the GPU frequency.}

\textcolor{black}{To validate our approach experimentally, we run
%
custom applications and commonly used graphics benchmarks on three different hardware platforms\footnote{
\textcolor{black}{Note that our previous work~\cite{Gupta2016Adaptive} was validated only on the Intel Minnowboard MAX mobile platform.}}:
the Intel Minnowboard MAX mobile platform~\cite{IntelMinnowboard},
Intel core i5 6$^{th}$ generation platform~\cite{paterna2017adaptive}, and Moto-X pure edition smartphone~\cite{motoX}}.
\textcolor{black}{First, we test the accuracy of our performance model.}
Our experiments show that the mean absolute percentage error in frame time and
frame time sensitivity prediction are \textcolor{black}{4.2\% and 6.7\%}, respectively.
\textcolor{black}{Then, we employ our model in a dynamic power management algorithm to
optimize energy consumption with performance constraint.
We achieve 43\% better energy savings than the default Ondemand governor
and only 3\% higher energy consumption compared to an Oracle policy.}

The major contributions of this work are:

\begin{itemize}
  \item A methodology for collecting offline data and developing a GPU performance model,
  \item \textcolor{black}{An adaptive performance model as a function of the GPU frequency and hardware counters
  observed online},
  \item \textcolor{black}{Practical implementation and overhead analysis of two low-cost RLS algorithms to adaptively learn the model coefficients},
  \item \textcolor{black}{Extensive evaluations of our approach on three experimental and commercial platforms using common GPU benchmarks.}
\end{itemize}

The rest of the paper is organized as follows.
Section~\ref{sec:related_work} presents the related work.
Section~\ref{sec:motivation} details the challenges and lays out the groundwork
required for frame time prediction.
Section~\ref{sec:methodology} presents the techniques for offline analysis and online learning.
Finally, Section~\ref{sec:experiments} discusses the experimental results,
and Section~\ref{sec:conclusion} concludes the paper.

%% file: related_research.tex
\section{Related Research} \label{sec:related_work}

\textcolor{black}{
The number of power hungry and performance critical graphics applications running on the smartphone \textcolor{black}{is} increasing~\cite{mochocki2006power}.
As a result, power consumption, temperature, and performance metrics in smartphones have become important \textcolor{black}{considerations}~\cite{gu2006games,Ogras2013Managing}.
Dynamic thermal and power management (DTPM) techniques often perform tradeoffs between these metrics for
good user experience\textcolor{black}{~\cite{Benini2000Survey, pothukuchi2016using, Gupta2016Generic}.}
This work focuses on building quantifiable light-weight performance models that can guide
DTPM algorithms in conjunction with runtime power models~\cite{jin2015towards, nagasaka2010statistical}.}

A number of researchers have proposed dynamic power management techniques for graphics applications~\cite{kadjo2015control, dietrich2010lms, Pathania2015Power}.
Many of these techniques employ performance models that are \textcolor{black}{either} learned offline or online.
For example, Kadjo et al. employs a performance model that is a function of the individual, the products, and the quotients of the
hardware performance counters~\cite{kadjo2015control}.
\textcolor{black}{This} technique learns the model parameters using batch linear regression and
predicts the frequency-scalable portion of the GPU active time.
\textcolor{black}{Thus, enabling} accurate performance modeling, but at the same time is dependent on the offline training data.
Another work on performance modeling uses an
auto-regressive (AR) model for frame time prediction~\cite{dietrich2014lightweight}.
The authors employ a tenth order AR model, whose coefficients are learned offline
using ten minutes of frame time data for each application using the
Matlab System Identification tool~\cite{MathworksMATLABb}.
In another technique, the authors use a similar AR model whose inputs are based on prior frame times,
and the model coefficients are estimated using the normalized least mean squares technique~\cite{dietrich2010lms}.

Workload prediction models based on PID controllers have also shown
good accuracy in prediction of graphics workloads~\cite{dietrich2010lms}.
However, as mentioned in~\cite{dietrich2010lms} the PID gains are very hard to tune due to
a large search space of the gain parameters.
Furthermore, it is not practically feasible to change the PID gains adaptively at runtime.
\textcolor{black}{Yet another} approach to compute the GPU performance is presented in~\cite{Pathania2015Power}.
This technique models the GPU performance using the CPU and GPU frequencies and their utilizations as inputs.
The authors employ batch linear regression adaptively at runtime to learn the model coefficients,
which is computation and memory intensive~\cite{sayed2003fundamentals}.
Furthermore, their model relies on utilization (instead of using the performance counters)
that does not provide a fine-grain measure of the workload.
%
%

\textcolor{black}{A hybrid combination of offline and online techniques has recently been proposed
to minimize the energy consumption under a performance constraint~\cite{mishra2015probabilistic}.
This technique employs probabilistic graphical models to estimate the power and performance for
unknown applications at runtime based on previously stored offline application data.
The authors show high accuracy compared to an online learning algorithm.
However, this online algorithm ignores the application history and employs a basic multi-variable linear regression technique.}

In summary, relying solely on offline data does not generalize well to other data sets,
as it is not feasible to account for all possible workloads.
Alternatively, online learning is challenging due to limited observability and computing resources.
We address these concerns by providing an efficient technique for GPU performance prediction, which
includes a performance model, a feature selection methodology and an online learning algorithm.

Our adaptive performance model uses hardware performance counters and frequency as inputs.
We employ RLS for online learning of the model coefficients.
Note that RLS has been extensively applied in signal processing and control applications~\cite{sayed2003fundamentals}.
In fact, RLS has also been employed for building an adaptive power model~\cite{wang2011adaptive}
and performance model~\cite{wang2016pareto, ma2014dppc} for CPUs.
Unlike our work, these models are not built for GPUs, and do not use frequency and performance counters as inputs.
Our prior performance model for integrated GPUs~\cite{Gupta2016Adaptive} also employ RLS algorithm and performance counters.
\textit{However, it \textcolor{black}{requires offline learning to characterize the frequency dependence of the counters used by the RLS algorithm.}}
\textcolor{black}{More precisely, the prior technique learns a non-linear model offline
to compute the derivative of frequency dependent counters with respect to the GPU frequency.}
\textcolor{black}{Since offline learning limits the usability of the earlier model, we propose a fully online technique.
The main challenge is to identify which counters depend on the GPU frequency, and characterize this dependence without knowing micro-architectural details.
Our key insight is to find this information in the experimental data set, which is already used for feature selection.
We add a subtle term to the model template used in the feature selection step.
The new term enables us to choose only the counters that are not correlated with the frequency term.
This leads to a more robust and practical mechanism that employs only frequency dependent counters.}

\textcolor{black}{In addition, we present the results with a low complexity DCD-RLS algorithm that can be more
efficient than traditional RLS algorithm for large number of inputs~\cite{Zakharov2008Low}.
\textcolor{black}{Furthermore, we also evaluate our technique by concurrently running GPU applications on
commercial Moto-X pure edition smartphone.
Finally, we demonstrate the application of our approach for dynamic power management and evaluate the results
on an Intel core i5 6$^{th}$ generation platform.}}

\begin{figure*}
	\centering
	\begin{subfigure}[b]{0.45\linewidth}
		\includegraphics[width=\linewidth]{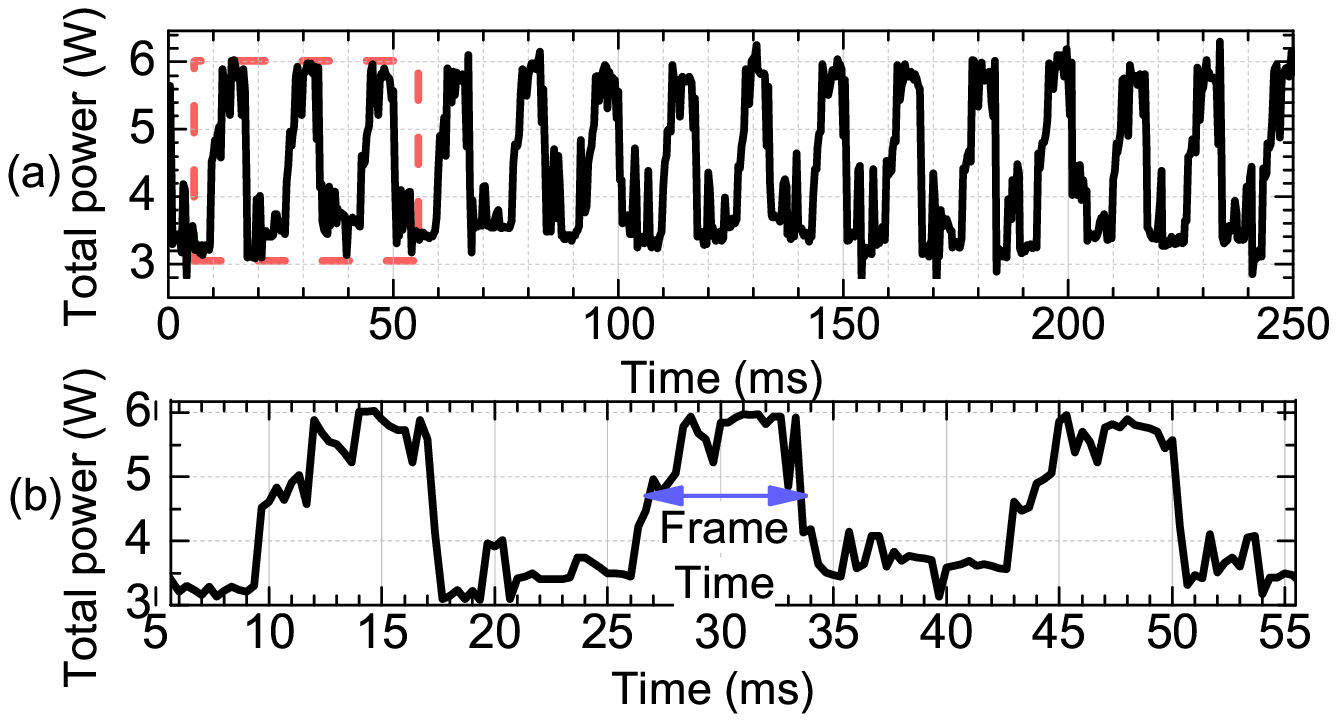}\\
		\label{fig:exp_setup_power_trace}
	\end{subfigure}
    \qquad
	\begin{subfigure}[b]{0.45\linewidth}
		\includegraphics[width=\linewidth]{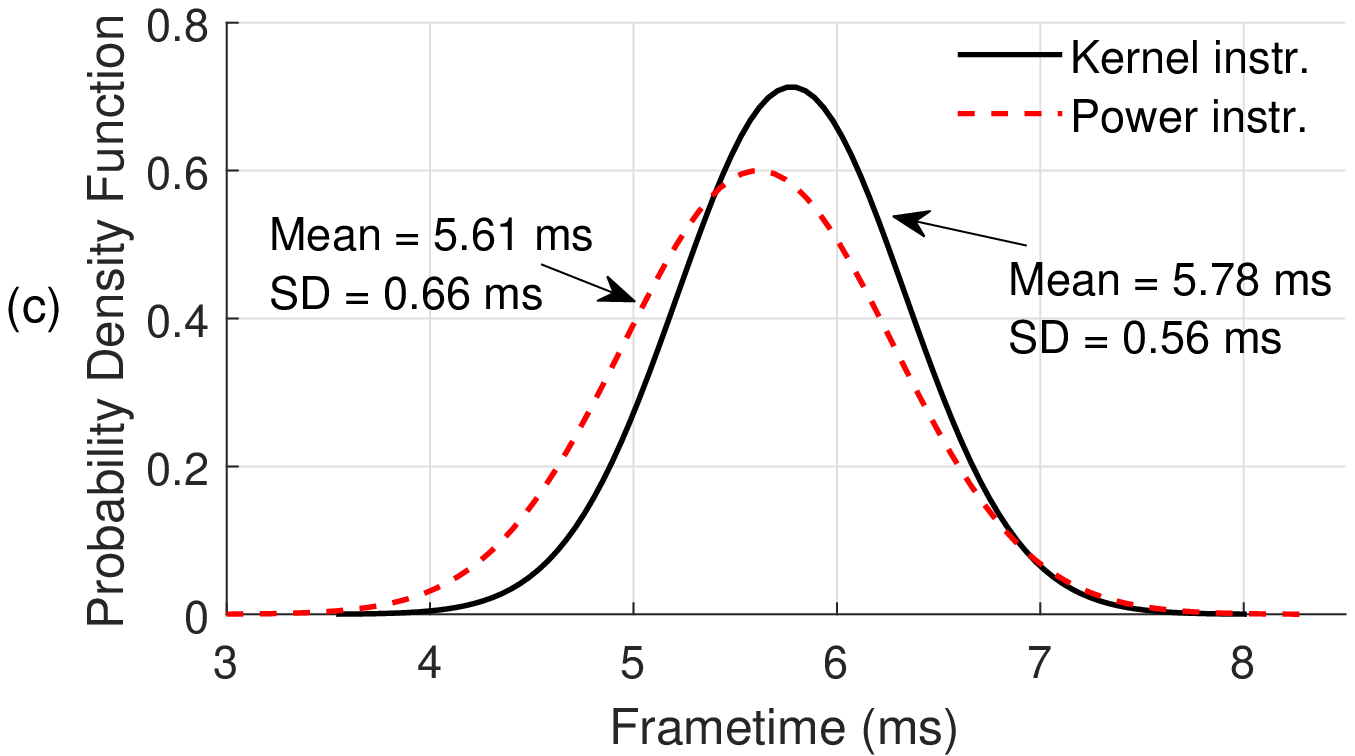}
		\label{fig:RLS_all_ave_trace_RenderingEngine_comp_actual_tF_dn_to_pred_tF_dn}
	\end{subfigure}
	\vspace{-2mm}
	\caption{(a) Total power consumption of the \textcolor{black}{Intel Minnowboard MAX platform~\cite{IntelMinnowboard}} when the GPU is rendering Art3 application at $60$ FPS. \textcolor{black}{The crests correspond to the power consumption when the GPU is actively rendering the frames, while the trough correspond to the power consumption when the GPU is in sleep state}. (b) Zoomed portion, which shows three frames in the first 50ms. \textcolor{black}{The width of the peaks give the time the GPU is actively computing the frame.}
		(c) Frame time distribution for kernel and power instrumentations for Art3 application.} \label{fig:frame_time_instr}
	\vspace{-2mm}
\end{figure*}

%% file: motivation.tex
\section{Frame Time Characterization}  \label{sec:motivation}

\subsection{Challenges and Notation} \label{sec:challenges}

\textcolor{black}{To construct a high fidelity frame time model, it is crucial to understand
the dependence of the frame time on the GPU frequency and workload.
As mentioned in Section~\ref{sec:intro}}, the workload characteristics are captured by the performance
counters $\mathbf{x} = [x_1, x_2, \ldots, x_N]$, where $N$ is the total number of counters.
\textcolor{black}{These counters are functions of the frame complexity
$C$ that can be defined as the processing effort required to render a frame.
For example, the number of various operations,
such as the number of pixels shades, and the number of cycles that the rendering engine is busy vary as the frames stream through the GPU.
Consequently, corresponding performance counters become indicators of the frame complexity.
Furthermore, both the frame time and some of the counters are functions of the operating frequency.
Therefore, we characterize the frame time $t_{F}$ in any given time step $k$
using a multivariate function $t_{F,k}(f_k,\mathbf{x}_k(C_k,f_k))$,
where the subscript $k$ denotes discrete time steps used in practical systems.}
Besides showing the dependency of the frame time on the frequency and counters,
this notation also reveals that the counters themselves can vary with frequency.

There are two major challenges in the characterization of $t_{F,k}(f_k,\mathbf{x}_k(C_k,f_k))$.
The first challenge is to establish a trusted reference \textcolor{black}{for frame time} that provides
a rich set of samples of this function.
This set needs to provide the frame time for an exhaustive list of frequencies and counter values.
The second and bigger challenge is to understand the sensitivity of frame time to frequency,
\emph{i.e.}, finding the partial derivative of the frame time with respect to the frequency.
\textcolor{black}{This quantitative measure of the impact on performance due to a change in the GPU frequency
is vital for dynamic power management algorithms.
For example, when the derivative is zero, the power management algorithm can safely lower the
frequency without affecting the performance.
However, finding this partial derivative is very challenging, since a direct reference is not
available at runtime.
Therefore, we perform extensive offline characterization by decoupling the impact of the change in frame time due to the frequency and frame complexity.}

\subsection{Frame Time and Counter Data Collection} \label{sec:data_collection}
\textcolor{black}{
We establish the reference frame time  by modifying
the Android's Direct Rendering Manager~\cite{faith1999direct} driver
to mark the GPU start and completion times for each new frame.
In this way, we can record the frame processing time and frame count
from the kernel while running any benchmark that uses the GPU.
We set the sampling period to 50~ms such that three frames
can fit into the interval at 60 frames per second (FPS).}

\textcolor{black}{Our frame time instrumentation is a non-trivial modification to the Linux kernel.}
\textcolor{black}{Therefore, we constructed an experimental setup to validate the accuracy of our instrumentation using power consumption measurements.
In our setup, an external power supply is connected to the target platform using a shunt resistor.
We employ an NI data acquisition (DAQ) system~\cite{NationalNI} to measure the voltage across the terminals of the resistor while running application.
Then, the data collected by the DAQ systems is used to compute the current drawn by the target platform.}
Figure~\ref{fig:frame_time_instr}(a) shows the total platform power consumption as a function of time
when running a custom target application (Art3) at 60 FPS.
By maintaining a low CPU activity, we know that the peaks in the power consumption occur due to the GPU activity.
Figure~\ref{fig:frame_time_instr}(b) zooms to the first 50~ms of the trace that shows three frames.
The width of the pulses is a good measure of frame time, since they correspond to the time periods during which the GPU is active.
Hence, we can test the accuracy of frame time and frame count instrumentations
by correlating them to the pulse durations obtained by power measurements.
\textcolor{black}{Figure~\ref{fig:frame_time_instr}(c) shows the probability density functions for the frame time measured by the software kernel instrumentation and the external board power measurements.}
We observe that our kernel instrumentation and power measurements yield only \textcolor{black}{a 3\% difference in mean of the frame time}.
We also find that the kernel instrumentation is more practical and accurate than the power measurements,
since it does not depend on external equipment and has lower measurement noise.

We use the Intel GPU Tools~\cite{IntelIntelb} to log the counter values at runtime~\cite{Intel2015Open}.
\textcolor{black}{Our modification to the kernel source code enables us to collect traces in the format shown below}:
\vspace{-0.02in}
\begin{table}[h]
\centering
\begin{tabular}{@{}llllllll@{}}
\toprule
\begin{tabular}[c]{@{}l@{}}Time \end{tabular} \hspace{-0.13in} & \begin{tabular}[c]{@{}l@{}}Frame \\ Time\end{tabular} \hspace{-0.13in} & \begin{tabular}[c]{@{}l@{}}Frame \\ Count\end{tabular} \hspace{-0.13in} & \begin{tabular}[c]{@{}l@{}}GPU \\ Frequency\end{tabular} \hspace{-0.13in} & \begin{tabular}[c]{@{}l@{}}Perf. \\ Cntr 1\end{tabular} \hspace{-0.13in} & \begin{tabular}[c]{@{}l@{}}Perf. \\ Cntr 2\end{tabular} \hspace{-0.13in} & $\dots$ \hspace{-0.13in} & \begin{tabular}[c]{@{}l@{}}Perf. \\ Cntr N\end{tabular} \\  \bottomrule
\end{tabular}
\end{table}
\\Each row corresponds to a 50~ms interval, which matches the rate at which the frequency governors change the GPU frequency.
We also test that this data collection does not induce any noticeable impact on the application performance.

\begin{figure}[t]
	\centering
	\vspace{-0.05in}
	\includegraphics[width=0.95\linewidth]{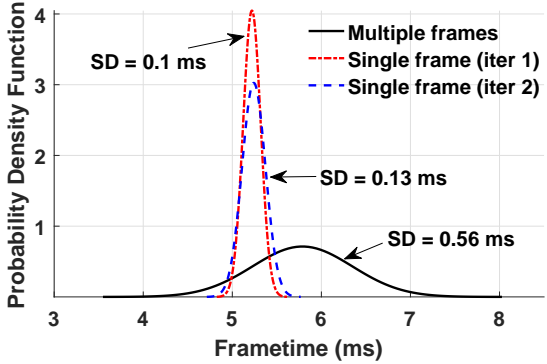}\\
    \vspace{-0.08in}
	\caption{The frame time distribution obtained for rendering the same frame and rendering multiple similar frames. 
    }
	\label{fig:exp_setup_repeated_frame_time}
\end{figure}

\textcolor{black}{Instead of collecting the data every 50~ms interval, another
way to isolate the changes due to the GPU frequency
is by running the entire application repeatedly at each supported GPU frequency.}
Theoretically, this \textcolor{black}{data collection method} can be
used to identify the effect of GPU frequency on frame time.
However, this approach is intractable for a number of reasons.
First, there will not be a one-to-one correspondence between the frames in different runs.
For example, consider an application that runs at 60 FPS or 30 FPS
depending on the GPU frequency.
At the low GPU frequency, the application will drop
the 30 frames that it failed to render, rather than rendering them later.
Second, even processing the same frame may take
different amounts of time due to the variations in the memory
access time from one run to another, as shown in Figure~\ref{fig:exp_setup_repeated_frame_time}.
\textcolor{black}{We also observe that frame time variations can be significant even when rendering multiple frames that
have similar frame complexity.}
These challenges are aggravated in many GPU intensive applications.
Therefore, the most reliable approach for collecting reference data is
by varying the GPU frequency while freezing the workload.

\begin{figure} [h]
	\centering
	\vspace{-0.11in}
	\includegraphics[width=0.85\linewidth]{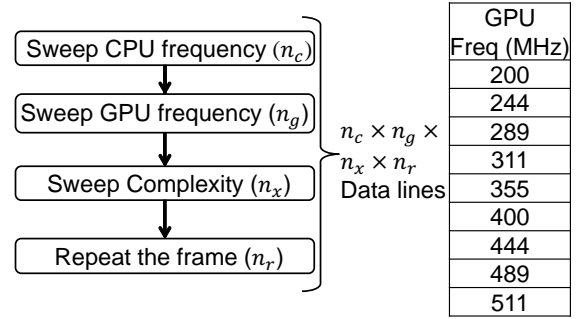}\\
     \vspace{-0.1in}
	\caption{The proposed methodology for collecting a rich set of training and test data.
Each frame is repeated $n_r$ times for every configuration.} \label{fig:data_set}
\vspace{-0.1in}
\end{figure}

\textcolor{black}{A consistent apple-to-apple comparison is possible
only if the workload is kept constant, \emph{i.e.}, same frame is frozen and rendered repeatedly.
To facilitate this comparison,} we built two custom Android applications,
Art3 and RenderingTest, as detailed in Section~\ref{sec:exp_setup}.
These applications enable us to precisely control the frame content and target frame rate.
\textcolor{black}{We first set the CPU frequency to ensure repeatability of the results, as shown in Figure~\ref{fig:data_set}.}
Then, we sweep the GPU frequency across the set of frequencies supported by the target system.
For example, our target platform supports nine frequencies ranging from 200MHz to 511MHz, as shown in Figure~\ref{fig:data_set}.
Each of these combinations is further repeated for 64 frame complexities,
which is determined by the number and variety of features in a given frame.
We note that different frame complexities enable us to exercise the performance counters in a controlled manner.
Finally, we run each configuration multiple times to suppress the random variations.
In our experiments, we collect 80 samples for each configuration,
which leads to 2 $\times$ 9 $\times$ 64 $\times$ 80 = 92160 lines
with 1152 different configurations.

\begin{figure} [t]
	\centering
	\includegraphics[width=\linewidth]{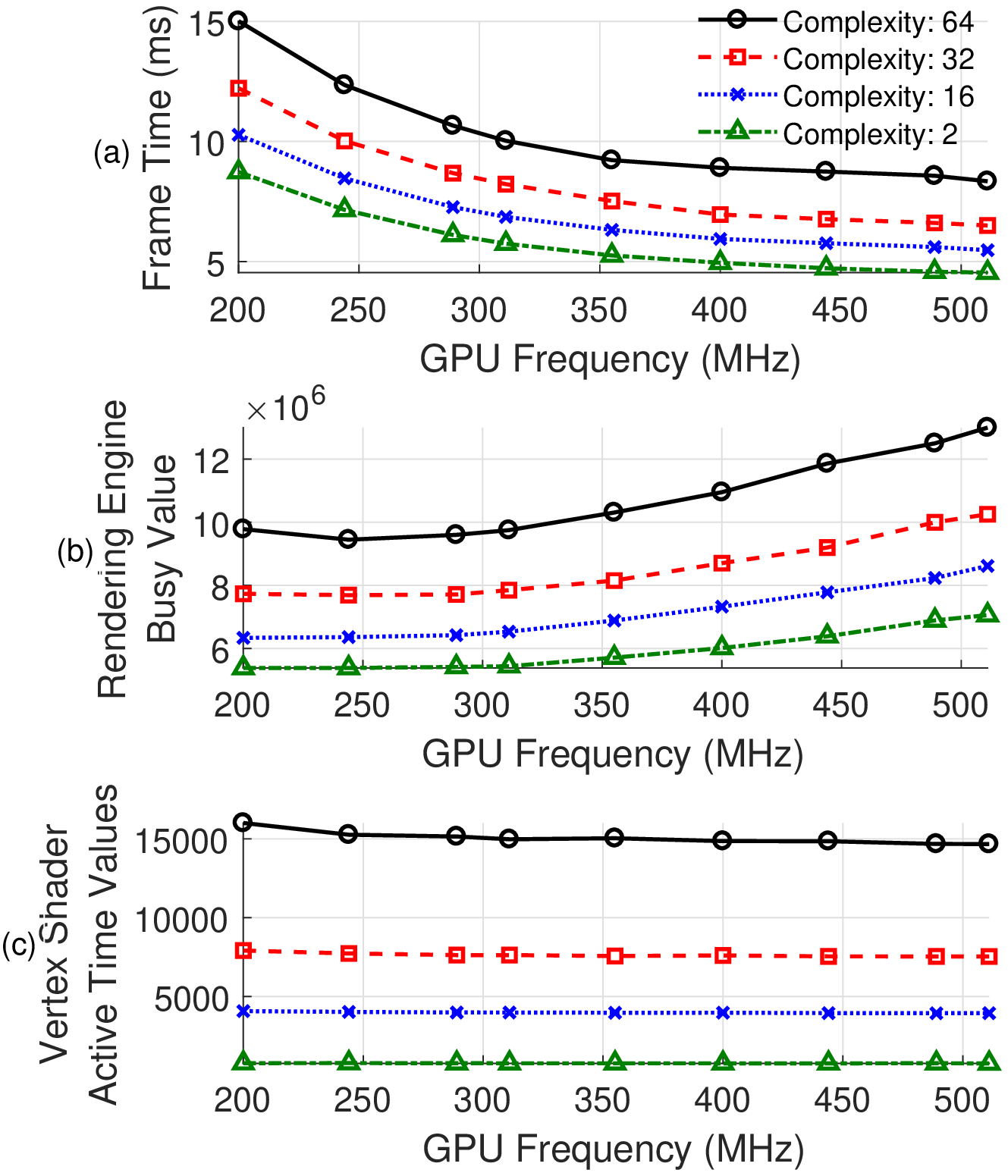}\\
	\caption{\textcolor{black}{Frame time and hardware counter values for the RenderingTest application with increasing GPU frequency at four different frame complexities.}}
	\label{fig:counter_comp_all_freq}

    \centering
	\vspace{2mm}
	\includegraphics[width=\linewidth]{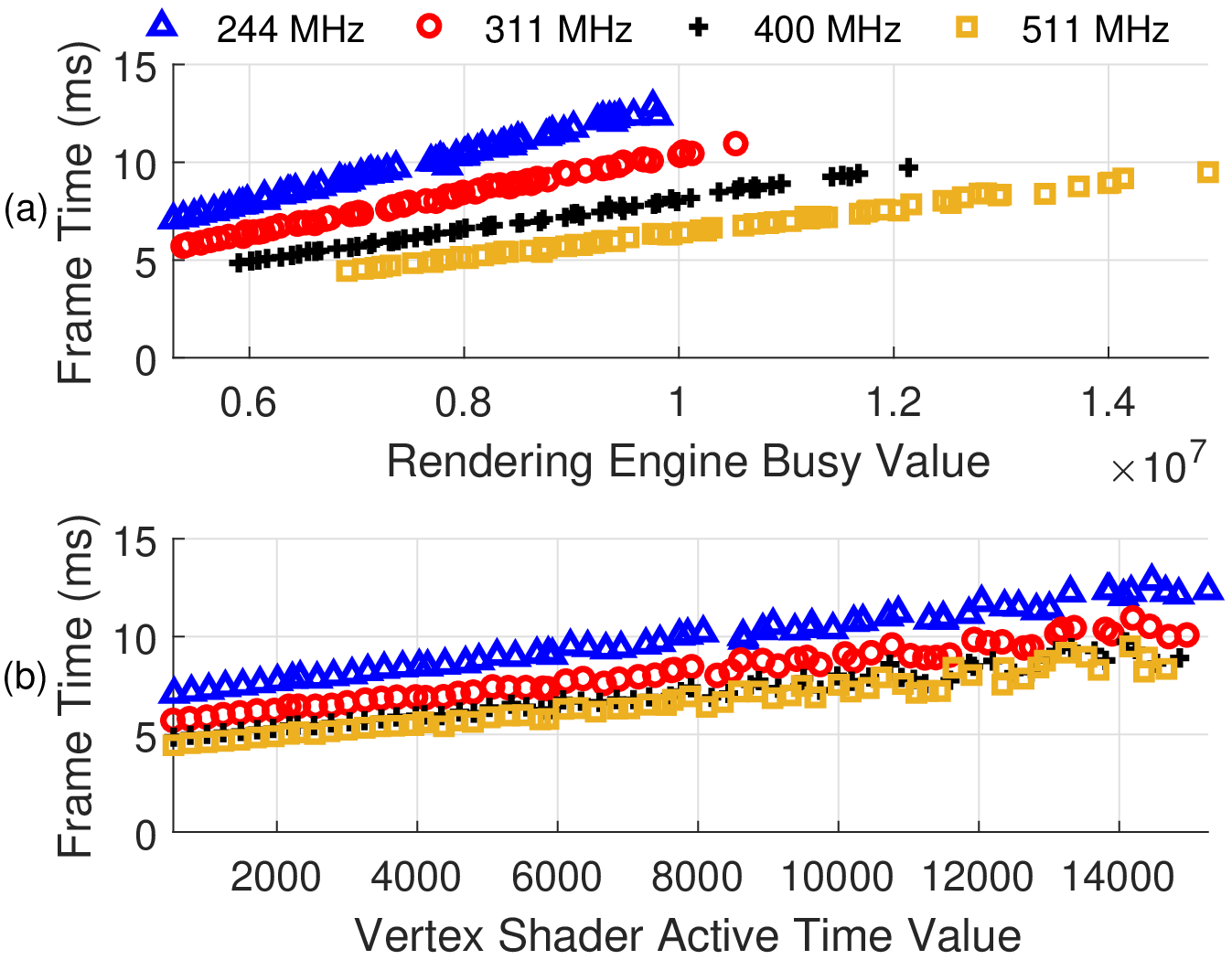}\\
	\caption{\textcolor{black}{Frame time for the RenderingTest application with increasing frame complexity at four different GPU frequencies.}}
	\label{fig:counter_freq_all_comp}

    \vspace{-5mm}
\end{figure}

\input{frametime_freq_counter_characterization.tex} 

%% file: frametime_freq_counter_characterization.tex
The proposed methodology is applied to both of our Art3 and RenderingTest applications.
\textcolor{black}{Figure~\ref{fig:counter_comp_all_freq}(a) shows how the frame time changes with the GPU frequency at a CPU frequency of 1.3~GHz.
Different curves on this plot show that increasing frame complexity implies larger frame time, as expected.
Therefore, the data set confirms that the frame time is a function of both the GPU frequency and the workload.}
\textcolor{black}{Similarly, Figures~\ref{fig:counter_comp_all_freq}(b) and (c) show the
\emph{Rendering Engine Busy} counter and \emph{Vertex Shader Active Time} counter as a function of the frequency.
The \emph{Rendering Engine Busy} counts the number of cycles
for which the rendering engine is not idle and
the \emph{Vertex Shader Active Time} counts the cycles for which the
vertex shader is active on all cores~\cite{Intel2015Open}.
Clearly, \emph{Rendering Engine Busy} counter is a strong function of frequency,
while \emph{Vertex Shader Active Time} counter is independent of frequency.
Figure~\ref{fig:counter_freq_all_comp} shows the relation between the
counters and the frame time.
We observe that a larger cycle count (\emph{i.e.,} higher complexity) results in an almost linear increase in frame time.
The partial derivative of frame time with respect to the counter value changes with frequency.
Furthermore, Figures~\ref{fig:counter_freq_all_comp}(a) and (b) show that the
partial derivative of frame time with respect to the counter value, \emph{i.e.}, the slope of the frame time,
is a function of both the frequency and counter.}
In summary, our data set enables characterizing the multivariate function $t_{F,k}(f_k,\mathbf{x}_k(f_k))$.
We use this data at design time to construct a template for the frame time model.
Then, our online learning algorithm updates the coefficients in this model to predict the frame time for arbitrary applications.

%% file: methodology.tex
\section{Frame Time Prediction} \label{sec:methodology}


This section presents the proposed frame time prediction methodology.
First, a mathematical model is derived to express change in frame time,
followed by a demonstration of how frame time sensitivity is computed using this model.
Then, we describe the offline learning process for selecting
the features that will be used during online learning.
Finally, we present the proposed adaptive frame time prediction algorithm.

\vspace{-0.01in}
\subsection{Differential Frame Time Model} \label{sec:frame_time_model}

\textcolor{black}{
The quintessential information used by dynamic power management algorithm is: ``How do the control parameters
(in our case the GPU frequency) affect the performance and power consumption''.
For example, if the performance is not affected by the GPU frequency, then we can use the minimum available frequency to minimize the power consumption,
since there is no performance penalty.
In contrast, if the frame time is inversely proportional to the GPU frequency, then it would be prohibitive to reduce the frequency.
Therefore, we are interested in modeling the change in frame time as a function of the frequency.
From a practical point of view, we know the frame time in the previous interval $k-1$ thanks to our instrumentation.
Therefore, the expected change (i.e., the difference from the previous interval) is sufficient to predict the frame time in next interval $k$.}
\textcolor{black}{This change can be approximated as the total derivative with respect to the GPU frequency and performance counters as follows:}

\small
\begingroup\abovedisplayskip=2pt \belowdisplayskip=3pt
\begin{multline}\label{eq:partial_derivative_active_time}
  dt_{F}(f_k,\mathbf{x}_k(C_k,f_k)) = \frac{\partial t_{F}(f_k,\mathbf{x}_k(C_k,f_k)) }{\partial f_k} df_k \\
  + \sum_{i=1}^N \frac{\partial t_{F}(f_k,\mathbf{x}_k(C_k,f_k))}{\partial {x_{i}}(C_k,f_k)} d{x_{i,k}}(C_k,f_k)
  \vspace{-0.02in}
\end{multline}
\endgroup
\normalsize

This equation reveals that the variation in frame time is a combined effect of
the change in the GPU frequency (the first term),
and the changes in the counters, which reflect the workload (the summation term).
Equation~\ref{eq:partial_derivative_active_time} holds,
if the frequency and counters are continuous variables.
Since they are discrete variables in practice,
we can approximate the change in frame time as:

\vspace{-0.05in}
\begin{equation}\label{eq:discrete_frame_time}
  \Delta t_{F}(f_k,\mathbf{x}_k(C_k, f_k)) \approx   \frac{\partial t_{F,k}}{\partial f_k} \Delta f_k + \sum_{i=1}^N \frac{\partial t_{F,k}}{\partial {x_{i,k}}} \Delta {x_{i,k}}
\end{equation}

Note that $\partial t_{F} / \partial f_k$ is the partial derivative of frame time with respect to frequency\footnote{
\textcolor{black}{We illustrate our approach using a single clock domain, since integrated GPUs used in mobile processors,
such as, ARM Mali, have a single clock domain~\cite{AnandTech_ARM_MALI}.
However, this approach can be extended to multiple clock domains by adding a new frequency term for each clock domain.
and using counters representative of all domains.}}.
The frame time change due to $\partial x_{i,k}(f_k) / \partial f_k$ is included in the difference term $\Delta{x_{i,k}}$.
This equation forms the basis of our mathematical model.
The differential form is useful, since the current frame time is known,
and we are interested in the change.
Next, we analyze each term of Equation~\ref{eq:discrete_frame_time} in detail to derive our frame time model.

\noindent \textbf{Change due to the GPU frequency:}
In general, the part of the processing time confined within the GPU pipeline
is inversely proportional to the frequency.
However, memory access and stall times do not scale with the frequency.
Therefore, the frame time is a non-linear function of the GPU frequency,
as shown in Figure~\ref{fig:counter_comp_all_freq}(a).
Using this observation, we can approximate the frame time $t_F$ \textit{for a given workload}
(\textit{i.e.}, $\mathbf{x}$)
in terms of a frequency-scalable portion $t_{F,s}$ and an unscalable portion $t_{F,us}$~\cite{ayoub2011OS}.
More specifically,

\vspace{-0.05in}
\begingroup\abovedisplayskip=2pt \belowdisplayskip=3pt
\begin{align} \label{eq:active_time_basic_model}
\begin{split}
  t_{F}(f_{k-1},\mathbf{x})  & = t_{F,s}(f_{k-1},\mathbf{x}) + t_{F,us}(\mathbf{x}) \\
  t_{F}(f_k,\mathbf{x}) & = t_{F,s}(f_{k-1},\mathbf{x}) \frac{f_{k-1}}{f_k} + t_{F,us}(\mathbf{x})
\end{split}
\end{align}
\endgroup
Hence, the change in frame time when switching from $f_{k-1}$ to $f_k$ can be found
by subtracting the first line in Equation~\ref{eq:active_time_basic_model}
from the second line as follows:

\small
\begin{equation}\label{eq:num_deriv_active_time_wrt_freq}
  \frac{\partial t_{F,k}}{\partial f_k} \hspace{-0.02in} \Delta f_k  \hspace{-0.02in} \approx \hspace{-0.02in}
  t_{F,s}(f_{k-1},\mathbf{x}) \hspace{-0.02in} \left(\hspace{-0.02in} \frac{f_{k-1}}{f_k}\hspace{-0.02in} - 1 \hspace{-0.02in} \right) \hspace{-0.02in} \equiv \hspace{-0.02in} a_0 t_{F,k-1} \hspace{-0.02in} \hspace{-0.02in} \left(\hspace{-0.02in} \frac{f_{k-1}}{f_k} \hspace{-0.02in} - 1 \hspace{-0.02in} \right)
\end{equation}
\normalsize
where $t_{F,k-1}$ is the frame time from the previous instant $k-1$.
We note that $ t_{F,k-1} \left( \frac{f_{k-1}}{f_k} - 1 \right)$ can be easily calculated at run time.
Since the scalable frame time is in general not known, we express it as \textit{an unknown parameter $a_0$}
that our online learning algorithm will learn at runtime.

%

\noindent \textbf{Hardware performance counter change:}
The frame time changes linearly with many hardware performance counters,
such as the one shown in Figure~\ref{fig:counter_freq_all_comp}.
If any counters cause a non-linear change in frame time, they can be taken as piece-wise linear.
Thus, we express the second term in Equation~\ref{eq:discrete_frame_time},
\textit{i.e.,} the change in frame time with counters as:

\vspace{-0.15in}
\begingroup\abovedisplayskip=2pt \belowdisplayskip=3pt
\begin{equation}\label{eq:frame_time__change_wrt_counter}
\Delta t_F(\mathbf{x}_k) \approx \sum_{i=1}^N \frac{\partial t_{F,k}}{\partial {x_{i,k}}} \Delta{x_{i,k}} \equiv \sum_{i=1}^N a_i \Delta  {x_{i,k}}
\end{equation}
\endgroup
where $a_i$'s are the coefficients that change at runtime as a function of the workload.
Therefore, they are learned online.

By combining Equation~\ref{eq:num_deriv_active_time_wrt_freq} and
Equation~\ref{eq:frame_time__change_wrt_counter}, we can re-write our mathematical model in Equation~\ref{eq:discrete_frame_time} as:

\small
\begin{multline}\label{eq:general_model}
\Delta t_{F,k}(f_k,\mathbf{x}_k(f_k)) \hspace{-0.02in} \approx \hspace{-0.02in} a_0 t_{F,k-1} \hspace{-0.04in} \left (\hspace{-0.02in} \frac{f_{k-1}}{f_{k}} - 1 \hspace{-0.02in} \right ) \hspace{-0.02in} + \hspace{-0.02in} \sum_{i=1}^N \hspace{-0.02in} a_i \Delta  {x_{i,k}}(f_k)
\end{multline}
\normalsize

The terms $t_{F,k-1}\hspace{-0.03in} \left( \hspace{-0.03in} \frac{f_{k-1}}{f_k} - 1 \hspace{-0.03in} \right)$ and $\Delta  {x_{i,k}}(f_k)~\forall~i \in [0,N]$ form the feature set $\mathbf{h}_k$,
while the parameters $\mathbf{a} \in \mathbb{R}^{N+1}$ are learned online.
\textcolor{black}{The list all of the parameters with their description are shown in Table~\ref{tb:notation}}.

\input{notation_table.tex}

\input{frame_time_sensitivity_theory.tex}
\input{feature_selection_theory.tex}

\input{online_learning.tex}

%% file: notation_table.tex
\begin{table}[b]
\normalsize
\centering
\caption{\textcolor{black}{Summary of the notation used in this paper}}
\label{tb:notation}
{\color{black}\begin{tabular}{@{}ll@{}}
\toprule
Notation & Description \\ \midrule
$k$ & Discrete time sample\\
$f$ & GPU frequency \\
$C$ & Complexity of a frame\\
$\mathbf{x} = [x_1, \ldots, x_N]$ & A vector of $N$ hardware counters \\
$T$ & Total number of data samples \\
$t_F$ & Frame time\\
$t_{F,s}$ & Frequency-scalable portion of frame time \\
$t_{F,us}$ & Unscalable portion of frame time \\
$\mathbf{a}$ & Model parameters\\
$f_\mathrm{new}$ & A new candidate GPU frequency \\
$N_\mathrm{indep}$ & Number of frequency independent counters \\
$\eta$ & $\ell_1$ regularization parameter\\
$M$ & Number of features after offline selection\\
$\mathbf{G}$ & Adaptive gain of the RLS \\
$\mathbf{P}$ & Covariance matrix of the error in RLS \\
$\mathbf{h}$ & Input features \\
$\lambda$ & Forgetting factor \\
$\mathbf{a}_\mathrm{init}$ & Initial estimate of the model parameters\\
$\mu$ & $\ell_2$ regularization parameter for RLS \\
\bottomrule
\end{tabular}}
\end{table} 

%% file: frame_time_sensitivity_theory.tex
\subsection{Frame Time Sensitivity}\label{sec:frame_time_sensititvity}


DTPM algorithms often need to evaluate the impact of a frequency change
on performance before making any decision.
This information, together with power sensitivity to frequency,
can help DTPM algorithms to make better decisions.
This section explains how our frame time prediction model is used for
computing the frame time sensitivity.

As an example, consider a scenario where the GPU frequency at time $k$ is $f_k = 400$ MHz.
Suppose that a DTPM algorithm needs to predict the change in frame time when the
frequency goes from $f_k = 400$ MHz to a candidate frequency $f_\mathrm{new} = 444$ MHz.
Before finalizing this decision, we will need to evaluate the corresponding change
in frame time,  \textit{i.e.}, $t_F(f_\mathrm{new}) - t_F(f_k)$ 
using Equation~\ref{eq:general_model}.
In this equation, the frequency change affects the first term
$\left( \frac{400}{444} - 1 \right)$ and only the counters that are a function of the frequency.
To make the latter more explicit, we can write the change in counters due to the GPU frequency $f$
and the frame complexity $C$ as:

\vspace{-0.05in}
\begingroup\abovedisplayskip=2pt \belowdisplayskip=3pt
\begin{equation}\label{eq:change_in_counters}
  \Delta x_{i,k} \approx \frac{ \partial x_{i,k} }{\partial f}  \Delta f_k + \frac{ \partial x_{i,k} }{\partial C} \Delta C, ~for~~1 \leq i \leq N
\end{equation}
\endgroup
Since the frame time sensitivity is calculated for a given frame, the change in complexity $\Delta C = 0$, and Equation~\ref{eq:general_model} can be written as:

\vspace{-4mm}
\small
\begin{multline}\label{eq:Delta_frame_time_due_to_freq1}
 \hspace{-0.13in} t_F(f_\mathrm{new}) \hspace{-0.01in} - \hspace{-0.01in} t_F(f_k)
  \hspace{-0.02in} \approx \hspace{-0.02in}
  a_0 t_{F,k-1} \hspace{-0.02in} \left(\hspace{-0.02in} \frac{f_{k}}{f_\mathrm{new}} \hspace{-0.01in}-\hspace{-0.01in} 1 \hspace{-0.02in} \right)
   \hspace{-0.01in} +  \hspace{-0.01in}
  \sum_{i=1}^{N} \hspace{-0.01in} a_i \hspace{-0.01in}
  \left(\hspace{-0.01in} \frac{ \partial x_{i,k} }{\partial f}  (f_\mathrm{new} \hspace{-0.02in}-\hspace{-0.02in} f_{k})  \hspace{-0.03in} \right)
\end{multline}
\normalsize

\textcolor{black}{This equation can be used to predict the change in frame time for the new candidate frequency as:
\begin{align}\label{eq:frame_time_sensitivity}
  \frac{dt_F}{df}\big|_k  \approx  \frac{t_F(f_\mathrm{new}) - t_F(f_k)}{f_\mathrm{new} - f_k}
\end{align}}
\vspace{-2mm}

In Equation~\ref{eq:Delta_frame_time_due_to_freq1}, $f_k$, $f_\mathrm{new}$, and $a_i~\forall~i \in [0,N]$ are known at time step $k$.
The only unknown value is $\frac{\partial x_{i,k} }{\partial f}$, which is zero for frequency \textit{independent} counters.
\textcolor{black}{Note that our prior work employed a non-linear offline
model to compute $\frac{\partial x_{i,k} }{\partial f}$~\cite{Gupta2016Adaptive}.
It is possible to learn this model online as well by employing two parallel adaptive algorithms,
but that will incur more overhead.
Since it is desirable to keep the overhead of the implementation small,
we modify the model to \textit{use only the frequency independent counters},
as described in Section~\ref{sec:feature_selection}.
Selecting the counters for which $\frac{\partial x_{i,k} }{\partial f} = 0$
greatly simplifies the frequency sensitivity calculation.
In particular, a simplified form can be obtained after combining the Equations~\ref{eq:general_model} and~\ref{eq:change_in_counters}, as follows (derivation is presented in Appendix~1):}

\small
\begingroup\abovedisplayskip=2pt \belowdisplayskip=3pt
\textcolor{black}{\begin{equation}\label{eq:general_model_freq_indep}
\Delta t_{F,k}(f_k,\mathbf{x}_k(f_k)) \hspace{-0.02in} \approx
 \hspace{-0.02in} a_0 t_{F,k-1} \hspace{-0.04in} \left (\hspace{-0.02in} \frac{f_{k-1}}{f_{k}} - 1 \hspace{-0.02in} \right ) \hspace{-0.02in} + \hspace{-0.02in} a_1  \Delta f_k  + \hspace{-0.02in} \sum_{i=1}^{N_\mathrm{indep}} \hspace{-0.02in} a_{i+1} \Delta x_{i,k}
\end{equation}}
\endgroup
\normalsize
\vspace{-2mm}

\noindent \textcolor{black}{where $N_\mathrm{indep} \subseteq N$ is the number of frequency independent counters.
This step simplifies the calculation of $t_F(f_\mathrm{new}) - t_F(f_k)$ 
by making the partial derivative of the counters with respect to frequency equal to zero in Equation~\ref{eq:general_model_freq_indep}.}

\small
\begingroup\abovedisplayskip=2pt \belowdisplayskip=3pt
\textcolor{black}{\begin{multline}\label{eq:Delta_frame_time_due_to_freq}
 \hspace{-0.13in}  t_F(f_\mathrm{new}) - t_F(f_k)
 \hspace{-0.02in} \approx \hspace{-0.02in}
  a_0 t_{F,k-1} \hspace{-0.02in} \left(\hspace{-0.02in} \frac{f_{k}}{f_\mathrm{new}} \hspace{-0.01in}-\hspace{-0.01in} 1 \hspace{-0.02in} \right)
   \hspace{-0.01in} +  \hspace{-0.01in} a_1 (f_\mathrm{new} \hspace{-0.02in}-\hspace{-0.02in} f_{k})
\end{multline}}
\endgroup
\normalsize

\noindent \textcolor{black}{\noindent \textbf{Derivative at time $k$:}
We can compute the derivative of frame time with respect to frequency at time $k$ using
the average of the derivative to jump one level higher
and one level lower frequency. The one level higher and lower frequencies
correspond to the smallest possible change in the
frequency of the platform.}

\vspace{-0.05in}
\begingroup\abovedisplayskip=2pt \belowdisplayskip=3pt
\textcolor{black}{\begin{multline}\label{eq:derivative2}
  \frac{dt_F}{df}\Bigg|_k = \lim_{\Delta f \rightarrow 0} \frac{1}{2} \biggl [ \frac{t_{F}(f_k + \Delta f)  - t_{F}(f_k)}{\Delta f} \\
                                + \frac{t_{F}(f_k)  - t_{F}(f_k - \Delta f)}{\Delta f} \biggr ]
\end{multline}}
\endgroup

\vspace{-2mm}
\noindent
\textcolor{black}{where $\Delta f$ is the change in the frequency one level higher and lower to the frequency $f_k$.
Since the change in the frequency is in both the higher and lower directions, the weights are 0.5.
For some platforms, such as Minnowboard the frequency levels are not equally spaced.
For example, when $f_k = 489$~MHz the change to the frequency
one level higher is $\Delta f_1 = 511-489 = 22$~MHz and one level lower is $\Delta f_2 = 489-444 = 45$~MHz,
as shown in the frequency table of Figure~\ref{fig:data_set}.
To accurately predict the numerical derivative of frame time with respect to the frequency,
we employ a three point derivative of Lagrange's polynomial~\cite{strang2007computational, singh2009finite} as follows:}

\vspace{-0.05in}
\small
\begingroup\abovedisplayskip=2pt \belowdisplayskip=3pt
\textcolor{black}{\begin{multline}\label{eq:lag_derivative}
  \frac{dt_F}{df}  \Bigg|_k \hspace{-2.5mm} \approx \hspace{-1mm} \frac{ \Delta f_1^2 t_F(f_k  \hspace{-0.5mm} + \hspace{-0.5mm}  \Delta f_2)
   \hspace{-1mm}  +  \hspace{-1mm}  (\Delta f_2^2 \hspace{-0.5mm} - \hspace{-0.5mm} \Delta f_1^2) t_F(f_k)
    \hspace{-1mm} -  \hspace{-1mm} \Delta f_2^2 t_F(f_k \hspace{-0.5mm} - \hspace{-0.5mm} \Delta f_1)}{\Delta f_1 \Delta f_2 (\Delta f_1  + \Delta f_2)}
\end{multline}}
\endgroup
\normalsize

\textcolor{black}{Equation~\ref{eq:lag_derivative} simplifies to Equation~\ref{eq:derivative2} for equal spacing of frequencies, \emph{i.e.}, when $\Delta f_1 = \Delta f_2$. }

%% file: feature_selection_theory.tex
\subsection{Offline Feature Selection} \label{sec:feature_selection}

Real-time prediction requires an extremely efficient learning algorithm
to facilitate fast evaluation of a GPU frequency change.
One approach to reduce the overhead of regression is dimensionality reduction on the input data.
The goal of this approach is to reduce the complexity of the data and speed up computation,
while maintaining a good prediction accuracy.
In addition to algorithm efficiency, this can help remove the features that either add duplicate information to the output or do not change with our parameters.
\textcolor{black}{The main challenge here is to
\textit{identify which counters depend on the GPU frequency} and
\textit{characterize this dependence without knowing micro-architectural details}}.
%
%
%
%
\textcolor{black}{We note the Equation~\ref{eq:general_model_freq_indep} has two types of terms.
The first two terms with coefficients $a_0$ and $a_1$ are explicit functions of the frequency,
whereas the remaining terms are functions of the performance counters.
If the counters in our feature set are correlated with the frequency, RLS cannot reliably converge to optimal model coefficients
due to the multicollinearity phenomenon.
Therefore, we limit our feature set to the performance counters that are independent from the frequency.
We are able to differentiate frequency dependent and independent counters using our characterization data
without having access to the micro-architecture of the GPU.}
We employ Least Absolute Shrinkage and Selection Operator regression (Lasso) to
reduce the feature size in the model appropriately by selecting the most representative
set of features by minimizing the MSE with a bound on the $\ell_1$ norm of parameters $a_i$~\cite{friedman2001elements}.
The results from Lasso regression are highly sparse due to the $\ell_1$ nature of the bound.
That is, for $T$ samples the Lasso regression can be performed by minimizing the
MSE between the actual change in frame time $\Delta t_{F,k}$ and
using the estimate from Equation~\ref{eq:general_model_freq_indep} after adding a $\ell_1$ norm penalty as:

\small
\vspace{-0.15in}
\begingroup\abovedisplayskip=0pt \belowdisplayskip=3pt
\textcolor{black}{\begin{multline}\label{eq:LassoEquation}
  \hat{a} = \argmin_a \hspace{-0.01in}
  \sum^T_{k = 1} \hspace{-0.01in} \biggl[\hspace{-0.01in}
  \Delta t_{F,k} - a_0 \hspace{-0.01in} t_{F,k-1} \left( \frac{f_{k-1}}{f_k} - 1 \right) \\ \hspace{-0.01in}-  \hspace{-0.02in} a_1  \Delta f_k  + \hspace{-0.02in} \sum_{i=1}^{N_\mathrm{indep}} \hspace{-0.02in} a_{i+1} \Delta x_{i,k} \biggr]^2
  + \eta \sum_{j = 0}^{N_\mathrm{indep}} |a_j|
\end{multline}}
\endgroup
\normalsize

By increasing the value of $\eta$, less features can be selected at the expense of accuracy.
An acceptable loss in accuracy is within one standard error more than the minimum MSE.
Thus, during the learning phase we will regress on $M$ feature subset, where $M << N + 1$,
instead of $N+1$ features.
\textcolor{black}{Note that our approach relies on the availability of frequency independent features
in the platform.
Based on our experiments with Minnowboard~\cite{IntelMinnowboard} and Intel core i5 6$^{th}$ generation platform~\cite{paterna2017adaptive},
we have always been able to find frequency independent features.}

%% file: online_learning.tex
\subsection{Online Learning of the Model Parameters}  \label{sec:online_learning}
The parameters in Equation~\ref{eq:general_model_freq_indep} can be learned offline and then used at runtime.
However, it is hard to generalize offline learning to all possible applications that would be executed by the system.
Moreover, the workload can change as a function of user activity.
Therefore, the learning mechanism should not completely rely on offline learning.
We employ an adaptive algorithm to learn the parameters of the frame time model.
\textcolor{black}{In particular, we use the covariance form of RLS~\cite{sayed2003fundamentals}
and the Dichotomous Coordinate Descent form of RLS~\cite{Zakharov2008Low} estimation techniques, as described next.}

\begin{figure} [t]
	\centering
	\includegraphics[width=0.8\linewidth]{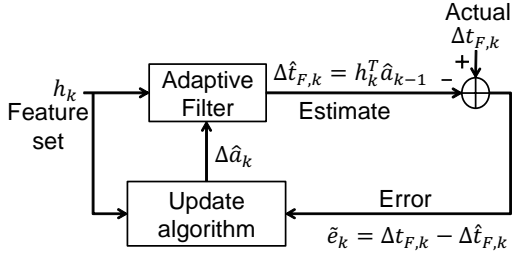}\\
	\vspace{-0.2in}
	\caption{Adaptive filtering approach showing the update in parameters $a_i$ based on error between the actual change in frame time and prediction.}
	\label{fig:rls_block_diagram}
	\vspace{-0.2in}
\end{figure}

RLS algorithm updates the parameters $a_i$ in Equation~\ref{eq:general_model_freq_indep}
in each prediction interval, as described in Figure~\ref{fig:rls_block_diagram},
using the following set of equations:
%
\begin{equation}\label{eq:RLS_gain_update}
\mathbf{G}_{k} = \mathbf{P}_{k-1}\mathbf{h}_k(\mathbf{h}_k^T\mathbf{P}_{k-1}\mathbf{h}_k + \lambda)^{-1}
\end{equation}
\begin{equation}\label{eq:RLS_projection}
\mathbf{P}_{k} = (\mathbf{I} - \mathbf{G}_k\mathbf{h}_k^T) \mathbf{P}_{k-1}\lambda^{-1}
\end{equation}
\begin{equation}\label{eq:RLS_paramter_update}
\hat{\mathbf{a}}_k = \hat{\mathbf{a}}_{k-1} + \mathbf{G}_{k} (\Delta t_{F,k}(f_k,\mathbf{x}_k(f_k)) - \mathbf{h}_k^T \hat{\mathbf{a}}_{k-1})
\end{equation}

 The update rule given in~Equation~\ref{eq:RLS_paramter_update}
computes the prediction error by subtracting the frame time prediction from the
actual change in frame time.
Note that online learning would not be possible without our kernel instrumentation, which provides
\textit{reliable reference measurement at runtime} ($\Delta t_{F,k}(f_k,\mathbf{x}_k(f_k))$).
Equation~\ref{eq:RLS_gain_update} and Equation~\ref{eq:RLS_projection} update
the gain $\mathbf{G}_k$ and covariance $\mathbf{P}_k$ matrices using the feature vector.
\textcolor{black}{The forgetting factor $0\ll\lambda\leq1$ is used to give more weight to
latest data and less weight to the older data. 
The set of Equations~\ref{eq:RLS_gain_update}-\ref{eq:RLS_paramter_update} together solve the
$\ell_2$ regularized cost function at runtime for any samples $T$ as follows~\cite{ismail1996equivalence}:}

\vspace{-2mm}
\begin{equation}\label{eq:cost_function_l2_regularization}
  \textcolor{black}{J = \min_{\mathbf{a}} \left [  \hspace{-0.1mm} (\mathbf{a} \hspace{-0.1mm}- \hspace{-0.1mm}\mathbf{a}_\mathrm{init})' (\mu \mathbf{I}) (\mathbf{a} \hspace{-0.1mm}- \hspace{-0.1mm}\mathbf{a}_\mathrm{init})
      \hspace{-0.3mm} +  \hspace{-0.3mm} \sum^{T}_{k=1} (\Delta t_{F,k} \hspace{-0.1mm} - \hspace{-0.1mm} \mathbf{h}_k'\mathbf{a})^2
      \hspace{-0.1mm} \right ]}
\end{equation}
%
\textcolor{black}{where $\mathbf{a}_\mathrm{init}$ are the initial values of the model coefficients $\mathbf{a}$ and $\mu$ is a regularization parameter. We denote the matrix and vector transpose by $(\cdot)'$ symbol.}


\noindent \textcolor{black}{\textbf{Parameter initializations:}
We choose the $\mathbf{a}_\mathrm{init} = diag(\mathbf{I})$, since we assume full scalability of the frames with respect to
the frequency and counters in the beginning.
The forgetting factor $\lambda$ is set to one to utilize all the past information.
We find the regularization parameter $\mu$ such that the multicollinearity of the inputs
is considerably reduced.
Multicollinearity in linear regression problems occur when two or more inputs are highly correlated
causing the standard errors in the estimate of the coefficients to increase~\cite{farrar1967multicollinearity}.
RLS solves the multicollinearity issue by minimizing a $\ell_2$ regularized cost function~\cite{hoerl1970ridge, ismail1996equivalence}.
Finally, we initialize the covariance matrix as $\mathbf{P} = \mathbf{I}/\mu $.}

\noindent \textbf{Computational complexity:}
RLS is well known for giving good predictions in the signal processing field. However, its computational complexity grows with the number of features as $O(M^2)$~\cite{sayed2011adaptive}.
Nonetheless, feature selection minimizes the size of the feature set to reduce the complexity. 
Furthermore, matrix inversions are the main source of complexity in many algorithms, including RLS.
Our solution is to use the co-variance form of RLS that does not perform matrix inversion.
The value $\mathbf{h}_k^T\mathbf{P}_{k-1}\mathbf{h}_k$ in Equation~\ref{eq:RLS_gain_update} evaluates to a scalar, eliminating the overhead of the inversion operation.
\textcolor{black}{The complexity of the RLS is acceptable for small number of features.
When there are large number of features then a traversal form of RLS coupled with
coordinate descent called DCD-RLS can be used~\cite{Zakharov2008Low}.
In this algorithm, first, the correlation matrix $\mathbf{P}^{-1}$ is partially updated in each time stamp $k$.
Then, the change in the model coefficients are estimated using inexact line-search.
This reduces the complexity of the DCD-RLS algorithm to O(M).
For example, in a platform if the number of features $M = 10$,
then the number of arithmetic operations in RLS are $2M^2 + 8M + 2 = 282$,
while the operations used in DCD-RLS are only $17M = 170$.
Since in our current platform we perform feature selection and reduce the number of
features to $4$, the number of operations in RLS and DCD-RLS are similar.
Also, DCD-RLS reduces the number of multiplication and division operations at the
expense of low-cost addition operations.
This provides slight speedup for small features and larger benefits when the number of features are more.
More details about the platform overhead of RLS are given in Section~\ref{sec:overhead_analysis}.}

%% file: experiments.tex
\section{Experimental Results}\label{sec:experiments}

%
%
%

\textcolor{black}{This section first describes the experimental setup and the selection of the offline learning of regularization parameters $\eta$ and $\mu$.
Next, we demonstrate the accuracy of the proposed online frame time and frequency sensitivity prediction techniques.
After that, we compare our approach to an existing online performance modeling methodology,
and demonstrate its application for dynamic power management.
Finally, we discuss the implementation overhead of the proposed frame time prediction techniques.}

\input{experimental_setup.tex}

\input{lasso_result.tex}

\input{result_freq_indep_cntrs.tex}

{\color{black}
\subsection{Comparison with an Auto Regressive Model using LMS}\label{sec:ar_lms}
\input{ar_lms.tex}

}

\begin{figure}[h]
	\centering	\includegraphics[width=\linewidth]{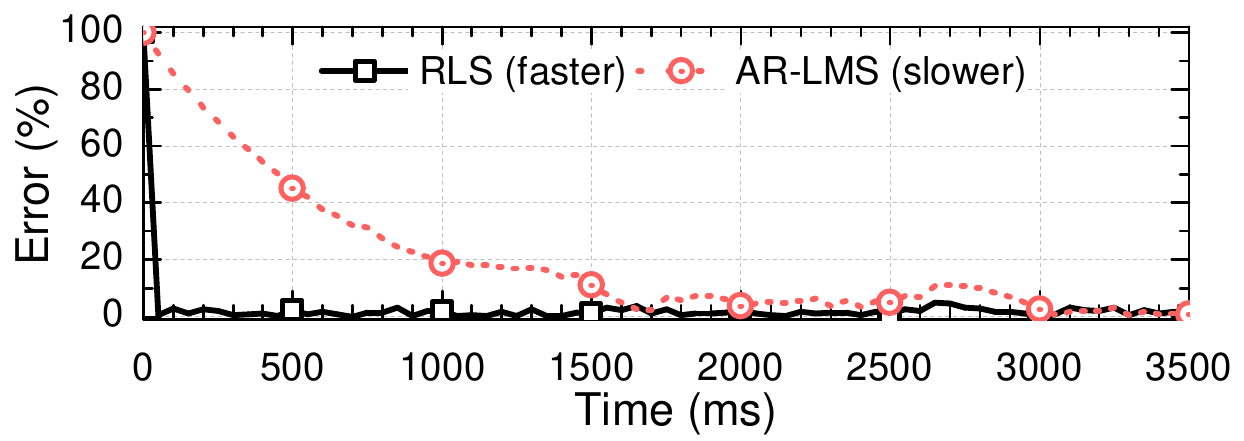}
	\caption{\textcolor{black}{The proposed RLS technique converges in only 50$ms$ compared to the AR-LMS technique that converges in 1.6$s$ for the Icestorm application.}}
    \vspace{-5mm}
	\label{fig:LMS_comp}
\end{figure}

{\color{black}
\subsection{Impact for Dynamic Power Management}\label{sec:dpm}
\input{dpm.tex}

}

\begin{figure}[t]
	\centering	\includegraphics[width=\linewidth]{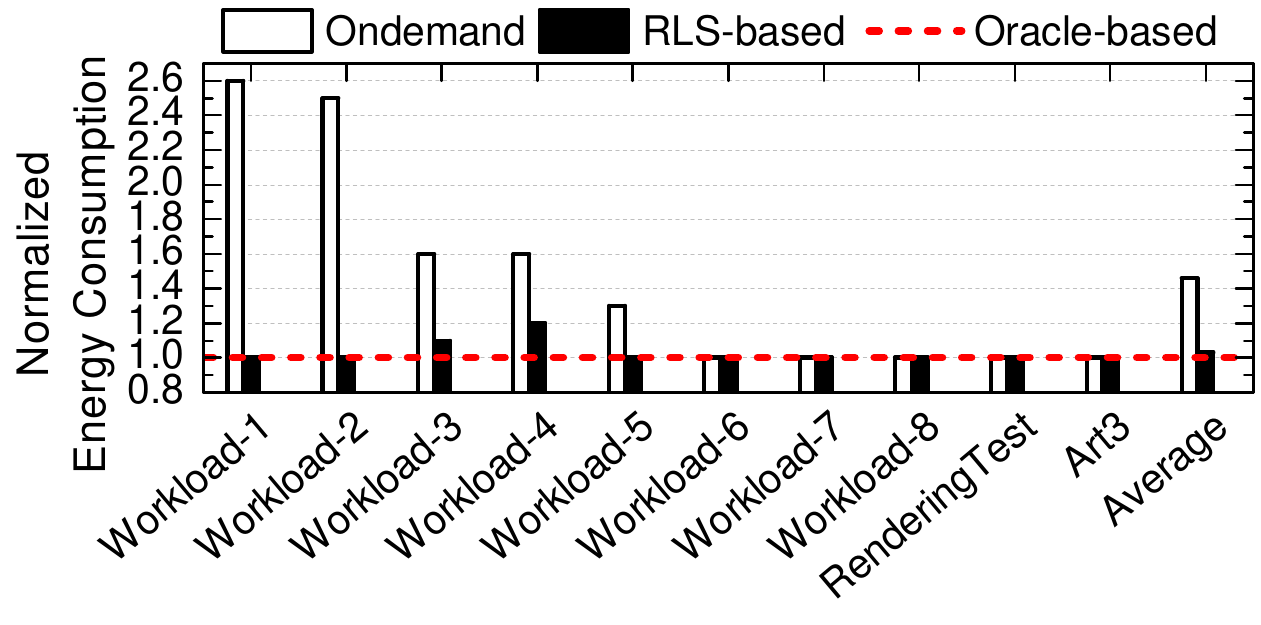}
	\caption{\textcolor{black}{
			Normalized energy consumption of the Ondemand governor and our RLS-based policy normalized to the Oracle-based policy.}}
	\label{fig:DPM_bar}
\end{figure}

\input{overhead.tex}

%% file: experimental_setup.tex
\subsection{Experimental Setup} \label{sec:exp_setup}

\textcolor{black}{We primarily employ the Minnowboard MAX platform~\cite{IntelMinnowboard} running the Android 5.1
operating system with the kernel modifications mentioned in Section~\ref{sec:data_collection} to evaluate our approach.}
This platform has two CPU cores and one GPU, whose frequency can take the values listed in Figure~\ref{fig:data_set}.
The GPU frequency is readily available from the kernel file system.
In addition to this, we use the Intel GPU Tools as an external module
to the Android system to trace the GPU performance counters.
%
%
\textcolor{black}{To further demonstrate the effectiveness of our approach, we employ two additional hardware platforms.
We evaluate the accuracy of our approach while running multiple graphics applications concurrently using
a Moto-X pure edition smartphone, which has Qualcomm Snapdragon 808 SoC.
Finally, we employ Intel core i5 6$^{th}$ generation platform~\cite{paterna2017adaptive}
for dynamic power management experiments.}

\noindent  \textbf{\textcolor{black}{Standard Benchmarks and Scenarios:} }
\textcolor{black}{The proposed frame time prediction technique is validated using the following commonly used GPU benchmarks on Minnowboard MAX platform:
Nenamark2, BrainItOn, 3DMark (both the Ice Storm and Slingshot scenarios), Mobilebench,
Chess, and Jet-ski.
We also employ eight gaming application scenarios, such as Fruit Ninja, Angry Birds,
Jungle Run, Angry Bots, and Shark Dash, running on Intel core i5 6$^{th}$ generation platform.
These workloads are referred to as Workloads 1-8 for confidentiality\footnote{This is requested by Intel Corp.}.
Finally, we run YouTube application and Chain Reaction game concurrently using Android 7 split-screen feature to create a multiple application scenario on Moto-X pure edition smartphone.}

\noindent  \textbf{Custom Benchmarks: }
The accuracy of the frame time prediction can be tested without any limitations,
since our frame time prediction technique works for any Android app that can run on the target platform.
However, validating the sensitivity prediction
(\emph{i.e.,} the partial derivative of the frame time with respect to the frequency)
requires reference measurements taken at different frequencies.
This golden reference cannot be simply collected by running the whole application
at different frequencies due to the reasons detailed in Section~\ref{sec:data_collection}.
Therefore, we also developed RenderingTest and Art3 applications
that enable us to control the number of times each frame is repeated.

The RenderingTest application accepts two inputs that specify the number of cubes rendered in the frame,
and the number of times the same frame is processed.
By changing the number of cubes, we control the frame complexity.
In our experiments, we sweep the number of cubes from 1 to 64
and repeat each frame 80 times.
The cubes are rendered at a maximum of 60 FPS with vertex shaders and depth buffering enabled.
Since we use the RenderingTest application for offline characterization,
we also developed one more custom application, called Art3, which renders pyramids
with a different rendering pipeline.
The RenderingTest application renders each cube with its own memory buffer,
while Art3 concatenates all pyramids into the same memory buffer before rendering.
These two applications allow us to compute and store the reference sensitivities,
such that they can be used as the golden reference to validate our online frequency sensitivity predictions.

\noindent  \textcolor{black}{\textbf{Evaluation:} We evaluate the proposed methodology using three algorithms.
The first algorithm employs Equation~\ref{eq:general_model_freq_indep} with online learning using RLS algorithm (RLS).
This is also the default algorithm used throughout the paper.
The second algorithm employs the same model with online learning using the DCD-RLS algorithm (DCD-RLS).
The third algorithm employs two models: (a) the model shown in Equation~\ref{eq:general_model} with online learning using RLS and
(b) an offline nonlinear model for derivative of frequency dependent counter with respect to frequency (RLS+Offline)~\cite{Gupta2016Adaptive}.
}

%% file: lasso_result.tex
\subsection{Offline Feature Selection and $\ell_2$ Regularization} \label{sec:exp_lasso}


\begin{figure}[t]
	\centering
	\includegraphics[width=\linewidth]{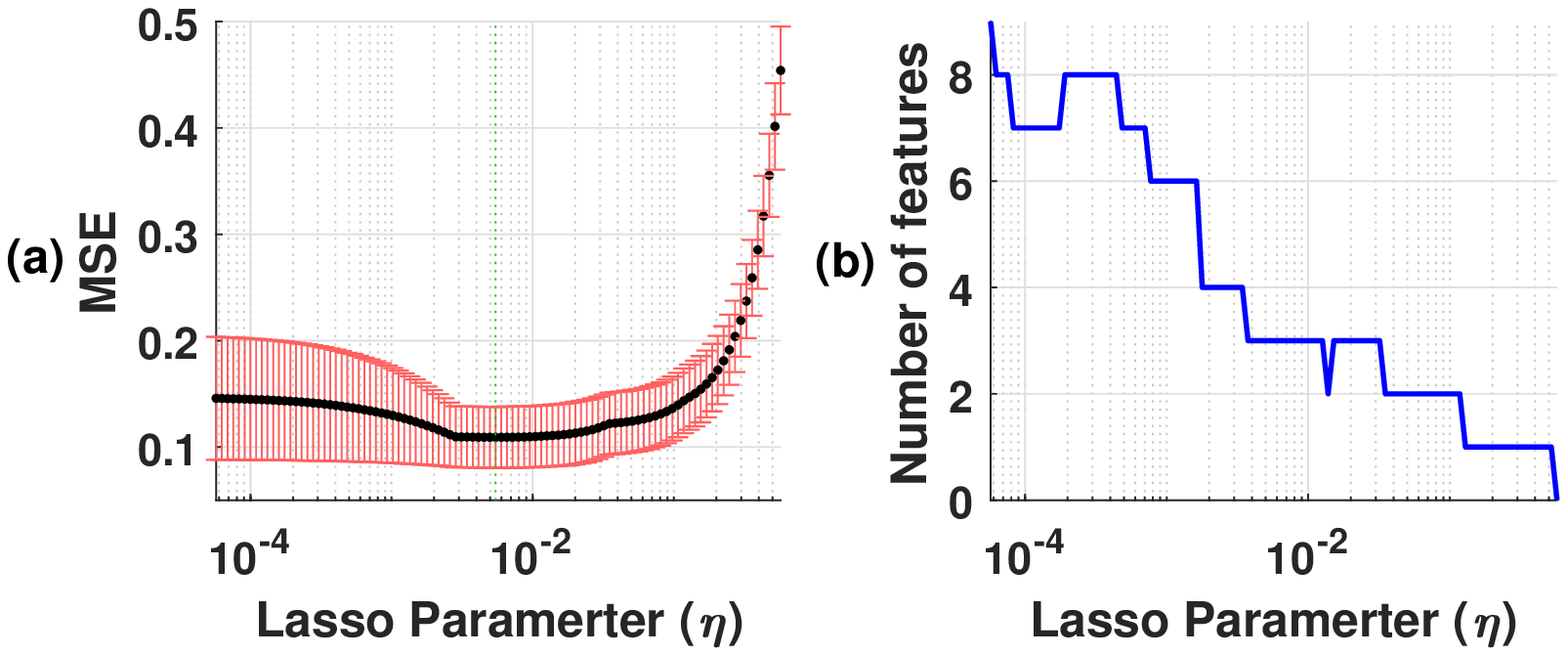}\\
\caption{
	 \textcolor{black}{Cross-validated LASSO regression result for; (a) the change in mean squared error of the frame time prediction with increasing $\eta$ values, and (b) the change in the number of selected features with increasing $\eta$ values.}}\label{fig:Lasso_result}
\end{figure}

To perform feature selection using Equation~\ref{eq:LassoEquation}, we first prune the counters that are highly dependent on frequency by
measuring the Pearson correlation coefficient of the counters with respect to the GPU frequency.
Counters that have the correlation coefficient less than 0.1 are retained for further processing.
Then, we apply the Lasso regression with 10--fold cross-validation
on our large dataset collected from the RenderingTest application.
Figure~\ref{fig:Lasso_result}(a) shows the change in mean squared error between the predicted and measured frame time of the GPU.
As the $\ell_1$ regularization parameter $\eta$ in Equation~\ref{eq:LassoEquation} increases,
the penalty on the cost function increases leading to a higher MSE, in general.
Note that the mean error (black line) first slightly decreases, then increases for incrementing $\eta$ values.
The slight decrease occurs due to overfitting that also leads to higher cross-validation variance in the error.
The minimum value of $\eta_\mathrm{min} = 5\times 10^{-3}$ uses four features, as shown in Figure~\ref{fig:Lasso_result}(b).
To shrink the model features, a good choice is $\eta_\mathrm{sel} = 3.4\times 10^{-3}$ for which the
performance in terms of expected generalization error is within one standard error of the minimum.
\input{exp_feat_selection_explaination.tex}
\textcolor{black}{\input{feat_corr_text.tex}}

\begin{figure}[h]
	\centering	\includegraphics[width=0.9\linewidth]{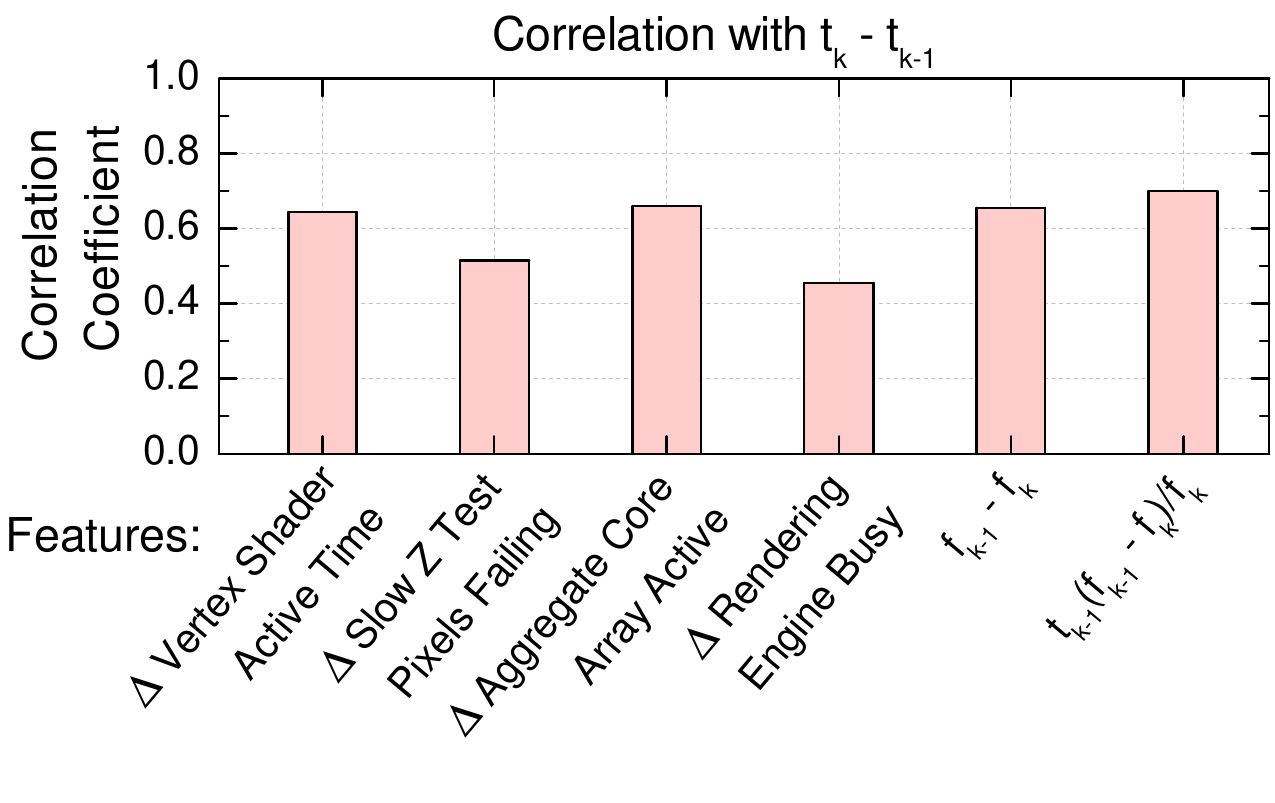}
\vspace{-5mm}
	\caption{\textcolor{black}{Correlation between the selected features and the difference in the frame time $t_k - t_{k-1}$.}}

	\label{fig:correlation_coeff_features_output}
\vspace{-2mm}
\end{figure}

\input{regularization_result.tex}

%% file: exp_feat_selection_explaination.tex
In our experiments, we choose the minimum MSE point with four features.
These four features consist of the two change in the frequency terms from Equation~\ref{eq:general_model_freq_indep}
and change in the \emph{Vertex Shader Active Time} and \emph{Slow Z Test Pixels Failing} counters.
The Vertex Shader Active Time counter counts the cycles for which the vertex shader is active on all cores.
The Slow Z Test Pixels Failing counter gives the number of pixels that fail the slow check in the GPU.
Neither of these counters depend on the frequency; they are functions of only the frame complexity.
Note that in our prior work~\cite{Gupta2016Adaptive} we also select four features, but these consist of a single
frequency change term and three counters.
Two of these counters, \emph{Aggregate Core Array Active} and \emph{Slow Z Test Pixels Failing} are frequency independent and one counter \emph{Rendering Engine Busy} is frequency dependent.
We compute the derivative of the frequency dependent counter offline using a non-linear model.
However, in this work by using frequency independent counters only, there is no need for using any additional models.
%
%
%
%

%% file: feat_corr_text.tex
Figure~\ref{fig:correlation_coeff_features_output} shows the features employed by our GPU performance model.
We observe that all the features are highly correlated to the change in frame time.

%% file: regularization_result.tex
\begin{figure}[t]
	\centering
	\includegraphics[width=\linewidth]{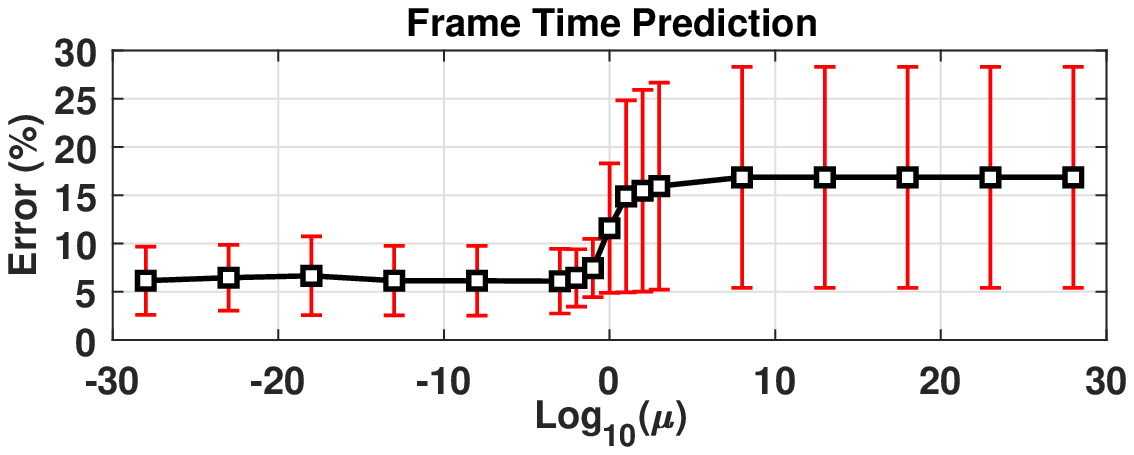}\\
\caption{
	 \textcolor{black}{Frame time prediction error for RenderingTest and Art3 applications for different values of the
$\ell_2$ regularization parameter $\mu$. The black markers show the mean value of the error and the whiskers show the one standard deviation boundaries.}}\label{fig:finding_mu}
    \vspace{-4mm}
\end{figure}

\textcolor{black}{We determine the $\ell_2$  regularization parameter $\mu$ for optimizing the cost function
in Equation~\ref{eq:cost_function_l2_regularization} of the RLS algorithm offline.
We first sweep the parameter $\mu$ between a large range of $10^{-28}$ to $10^{20}$,
and run the RLS algorithm for each value of the $\mu$ to find the
error in the frame time predictions.
Figure~\ref{fig:finding_mu} shows the mean and variance of the
absolute percentage error in frame time for a number of $\mu$ values for the RenderingTest and Art3 applications.
When $\mu$ is small, there is little regularization effect and consequently the error is low. 
However, when $\mu$ value is large, the left term in Equation~\ref{eq:cost_function_l2_regularization}
starts to dominate the cost function and severely constrains the model coefficients $\mathbf{a}$
close to $\mathbf{a}_\mathrm{init}$.
This leads to higher frame time prediction errors for $\mu > 1$.
We employ a $\mu = 10^{-14}$ in all our experiments, which is the geometric mean of the starting sweep value of $\mu = 10^{-28}$
and the knee point $\mu = 1$ to provide sufficient adaptability for any unknown workloads.}

%% file: result_freq_indep_cntrs.tex
\subsection{Online Frame Time Prediction} \label{sec:frame_time_prediction_results}

\begin{figure} [h]
\centering

    \includegraphics[width=\linewidth]{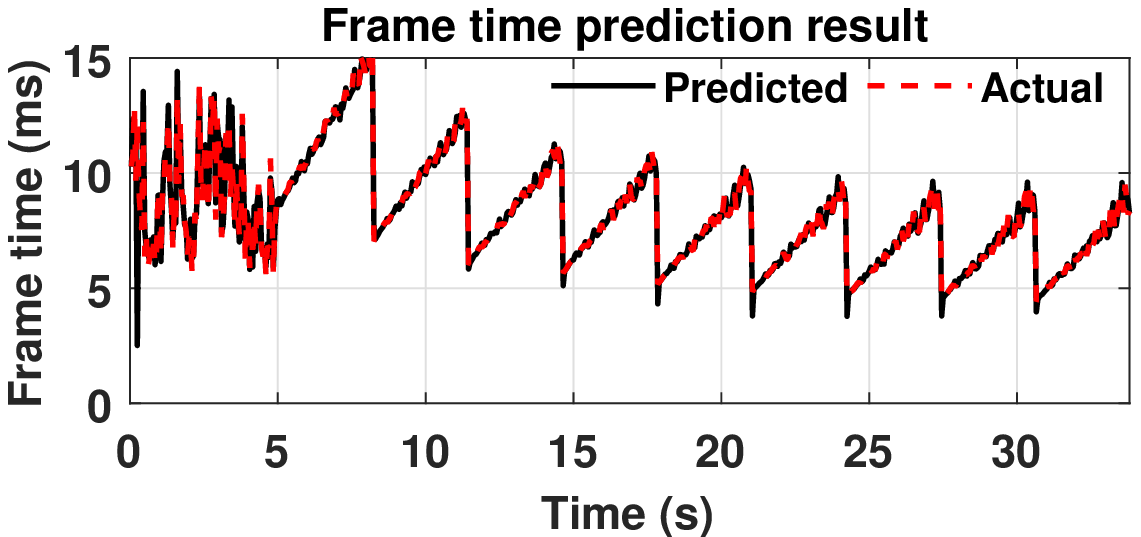}\\
    \caption{\textcolor{black}{Frame time prediction for the RenderingTest app.}}
        \label{fig:FrameTime_comparision_trace}
    \vspace{-2mm}
\end{figure}

We validate our frame time prediction approach first on the RenderingTest application to test the corner cases.
Figure~\ref{fig:FrameTime_comparision_trace} shows the comparison between the actual
and the predicted frame time.
During the first $5$ seconds, both the GPU frequency and frames change randomly.
We observe that the proposed online model successfully keeps up with the rapid changes.
In order to test our approach under corner cases, we enforce a saw-tooth pattern
during the remaining duration of the application.
More precisely, the GPU frequency starts at $200$ MHz, and
the complexity increases from $1$ to $64$ in increments of one (the first tooth).
Then, the same iterations are repeated for $9$ supported GPU frequencies.
Figure~\ref{fig:FrameTime_comparision_trace} demonstrates that we achieve
very good accuracy when the frequency stays constant for a period of time.
There is a spike when the complexity jumps suddenly from $64$ to $1$.
However, the RLS reacts quickly and maintains a high accuracy.
Overall, the mean absolute percentage error between the real and predicted frame time values is $2.6\%$.

\begin{figure} [h]
	\centering
	\includegraphics[width=\linewidth]{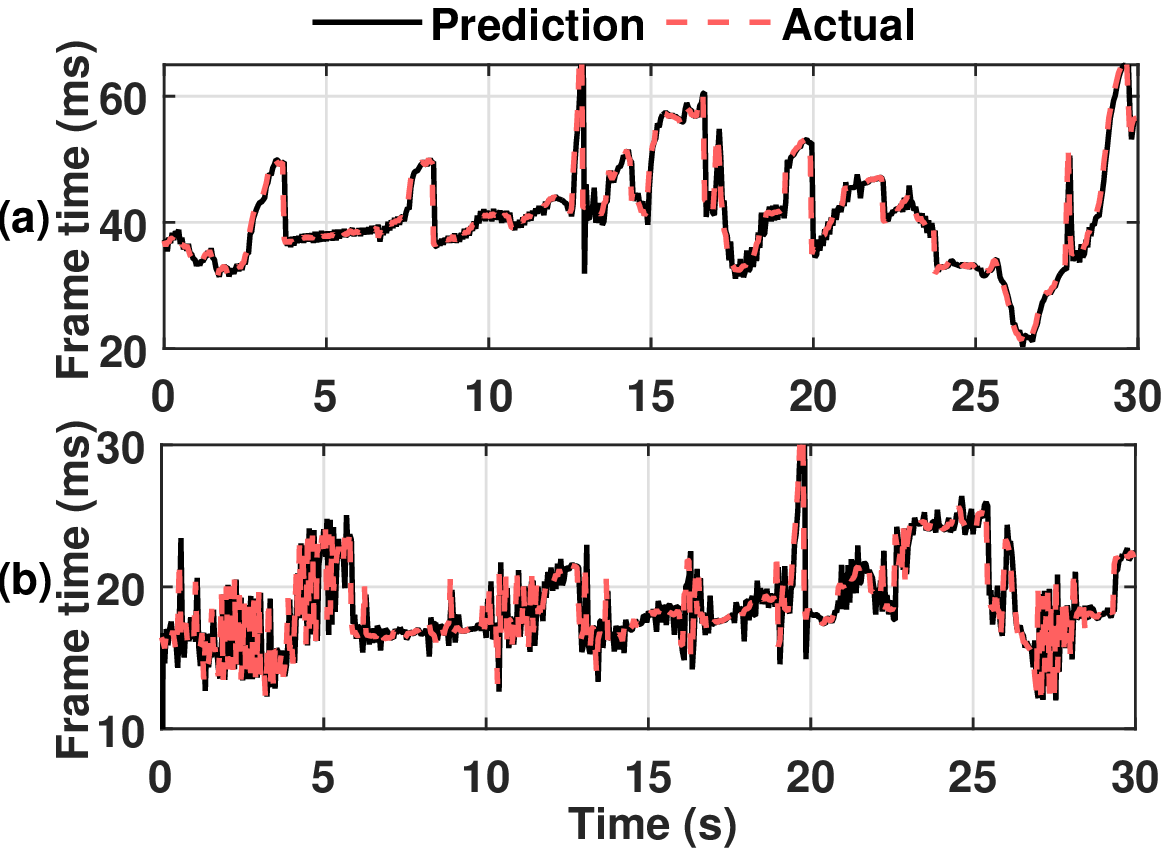}\\
	\caption{\textcolor{black}{Frame time prediction for the 3DMark Ice Storm application running at (a) $200$ MHz, (b) $489$ MHz.}}\label{fig:icestorm_experiment_trace_result}
    \vspace{-2mm}
\end{figure}

\begin{figure} [h]
	\centering
	\includegraphics[width=0.95\linewidth]{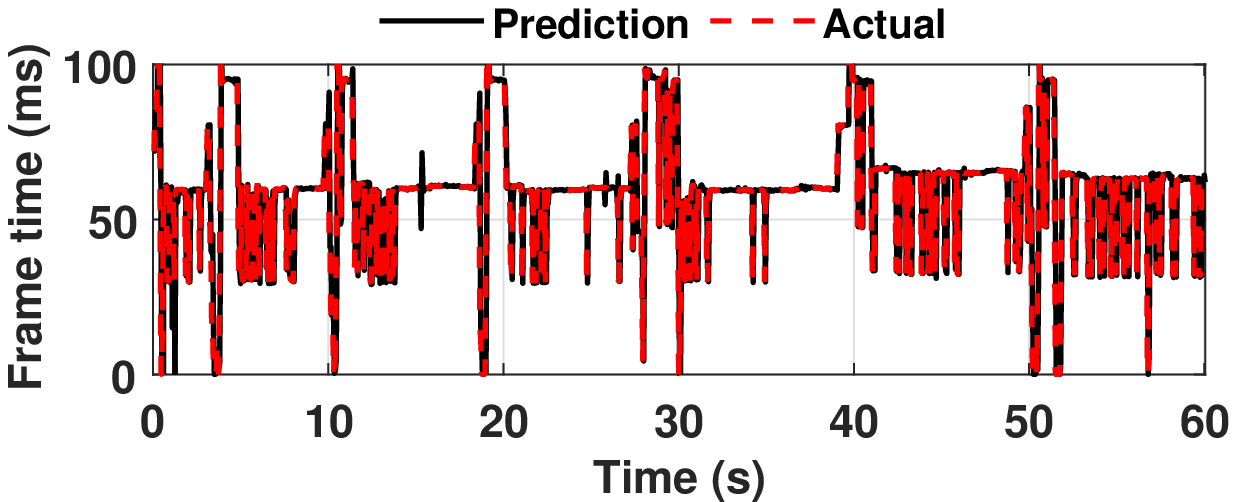}\\
	\caption{\textcolor{black}{Frame time prediction for the BrainItOn application running at $200$ MHz.}}\label{fig:BrainItOn_RLS_result}
    \vspace{-2mm}
\end{figure}

We observe similar levels of accuracy for Art3 and standard benchmarks.
In particular, Figure~\ref{fig:icestorm_experiment_trace_result} shows the actual and predicted frame times for 3DMark's Ice Storm benchmark at two different GPU frequencies.
We achieve a high prediction accuracy with the mean absolute error of \textcolor{black}{$2.1\%$ and $7.4\%$} for the GPU frequencies $200$ MHz and $489$ MHz, respectively.
Similarly, the actual and predicted frame time for the BrainItOn gaming application with fixed GPU frequency is shown in Figure~\ref{fig:BrainItOn_RLS_result}.
This interactive game requires frequent user inputs, and the frame time exhibits more sudden changes compared to other applications.
Our frame time prediction matches closely to the actual frame time with the median and mean absolute percentage errors of \textcolor{black}{$0.4\%$ and $12.9\%$}, respectively.
Note that the higher mean absolute error value for the application is due to a few outliers in the frame time.
This is confirmed from the very low median absolute percentage error value of the benchmark.

\begin{figure}[t]
\centering
    \includegraphics[width=\linewidth]{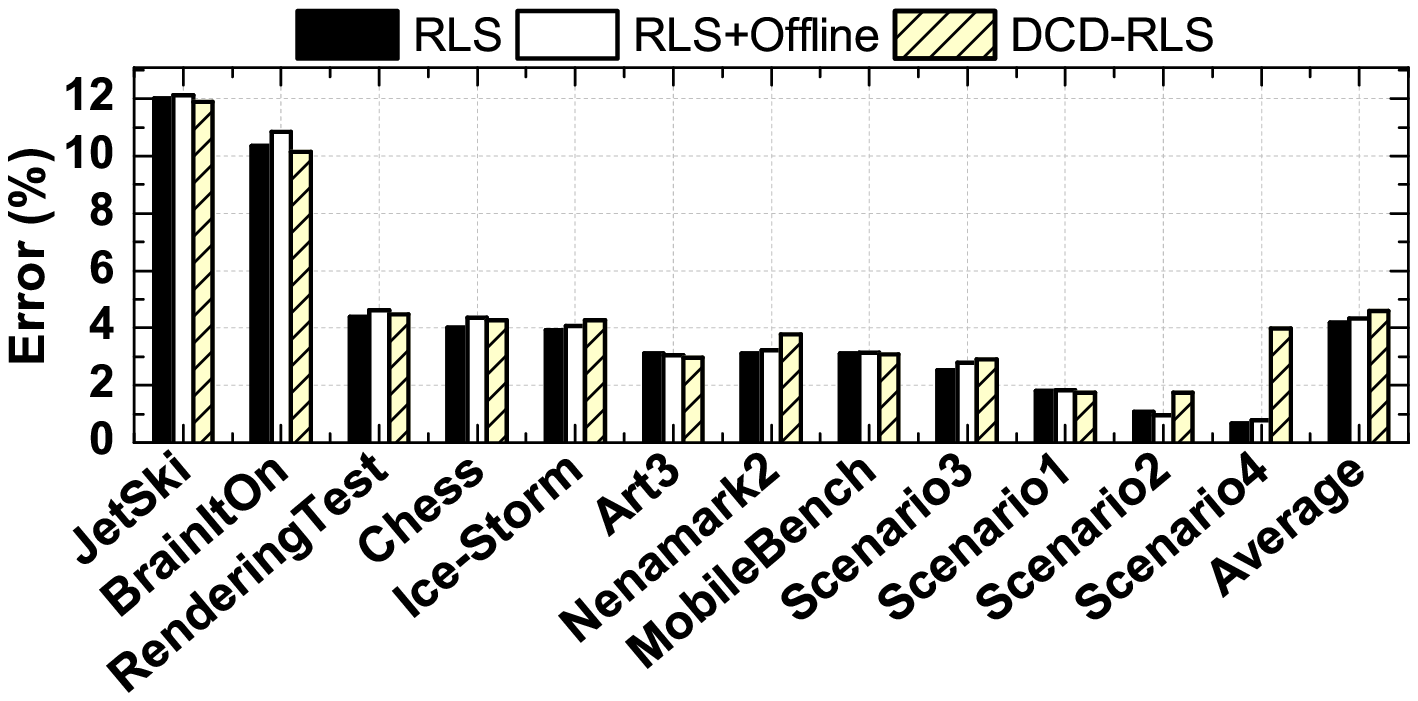}\\
    \caption{\textcolor{black}{Mean absolute percentage errors in the frame time for the Android applications using the three algorithms: RLS, RLS+Offline, and DCD-RLS.} }
        \label{fig:RLS_result_summary}
\end{figure}

The frame time prediction mean absolute error for all of the benchmarks running over \emph{all GPU frequencies} is summarized in Figure~\ref{fig:RLS_result_summary}.
The results are sorted with the errors in the RLS technique.
The average of the mean absolute errors across all the benchmarks for the RLS,
RLS+Offline, and DCD-RLS algorithms are $4.2\%$, $4.3\%$, and $4.6\%$, respectively.
On average, the three algorithms provide similar and high accuracy.
The RLS and DCD-RLS techniques have the additional advantage of not relying on any offline model.
\textcolor{black}{
\input{tferror_high_explaination.tex}}
%
%
%
%
%
%
Other benchmarks show errors smaller than $5\%$, indicating very high accuracy for frame time
prediction. For Scenario-4 benchmark the DCD-RLS technique shows $3\%$ higher error compared to the other two algorithms.
This is because the RLS algorithm is better at rejecting the noise in the inputs compared to the DCD-RLS.
This indicates that RLS should be preferred over DCD-RLS,
except when the complexity of RLS is critically important in the system
and slightly larger errors in frame time prediction are acceptable.

\begin{figure}[t]
\centering
	\includegraphics[width=0.9\linewidth]{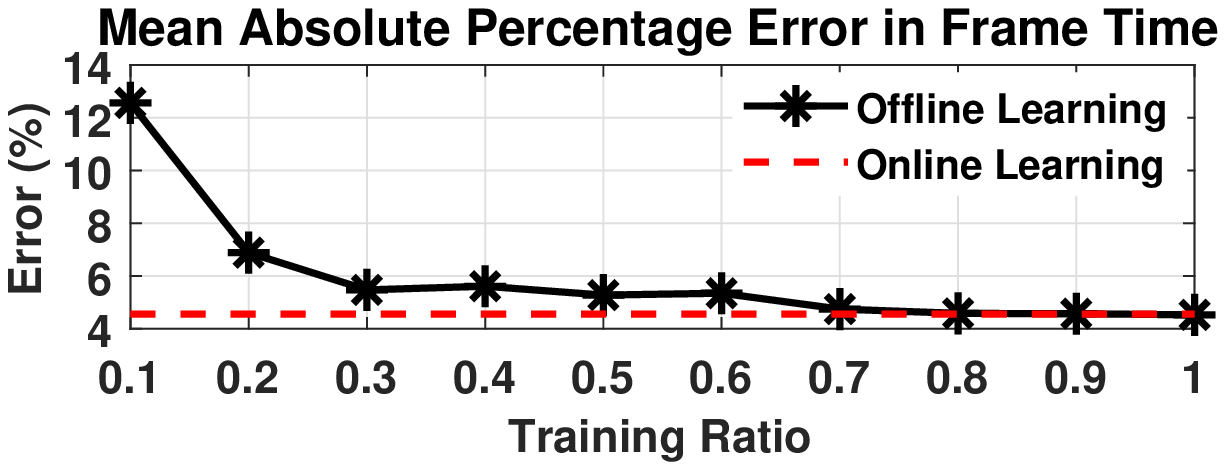}\\
	\caption{\textcolor{black}{Comparison of mean absolute percentage error in frame time for all Android applications combined.}}\label{fig:offline_vs_online}
\end{figure}

\noindent \textbf{Comparison with Completely Offline Learning:} We also compare our approach with an offline method,
where all the model parameters are learned at design time and remain constant at runtime.
Figure~\ref{fig:offline_vs_online} shows the mean absolute percentage errors for online (dashed line) and offline (solid line) learning for different training ratios.
When we run all the benchmarks one after the other with our online learning mechanism, we get an error of $4.6\%$.
However, running the same benchmarks with offline learned parameters leads to higher errors.
As shown in the figure, the difference between the offline and online error decreases as the training ratio approaches one, \emph{i.e.}, when the training set equals the test set.
This shows that offline learning leads to higher error, unless the model can be trained on all the applications.
Of note, the prediction error of our approach is flat, since the same set of features are selected with smaller training set.

{\color{black}
\input{multi_apps.tex}

}

{\color{black}
\input{time_method_motox.tex}
}

\begin{figure}[t]
	\centering	\includegraphics[width=\linewidth]{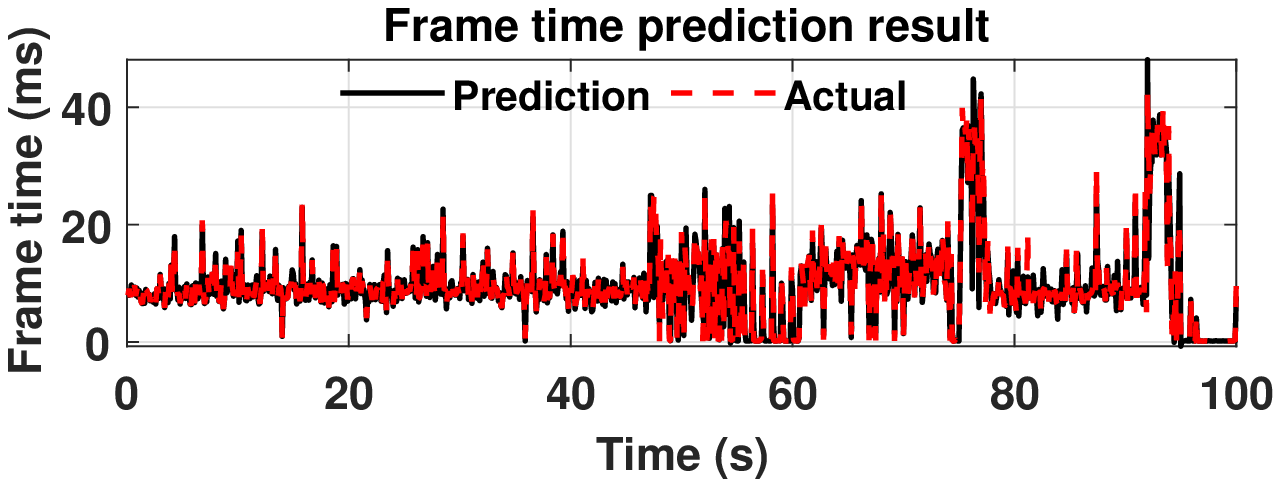}
	\caption{\textcolor{black}{Frame time prediction while running YouTube and Chain reaction game running simultaneously on Moto-X smartphone.}}
	\label{fig:MotoX_multiApps}
\end{figure}

\subsection{\textcolor{black}{Online Frame Time Sensitivity Prediction}} \label{sec:potential_impact_on_DPM}

\begin{figure}[t]
    \centering
    \includegraphics[width=0.9\linewidth]{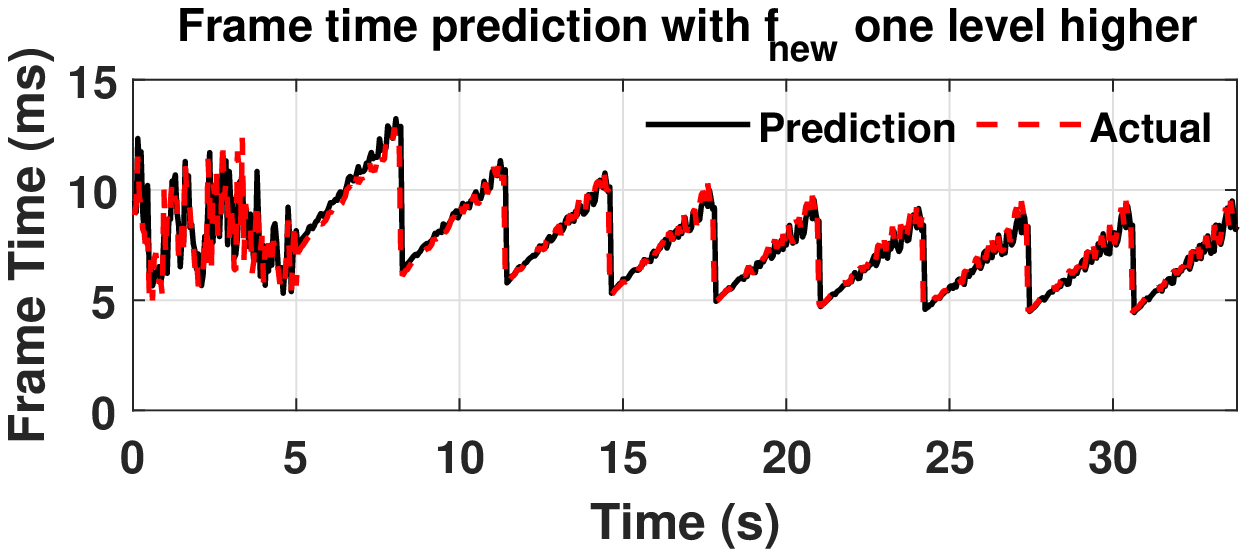}
    \caption{\textcolor{black}{Predicted and actual frame times for RenderingTest application when $f_\mathrm{new}$ is one level higher.}}
    \label{fig:RLS_all_ave_trace_RenderingEngine_comp_actual_tF_up_to_pred_tF_up}
\end{figure}

\begin{figure}[h]
	\centering
	\includegraphics[width=0.9\linewidth]{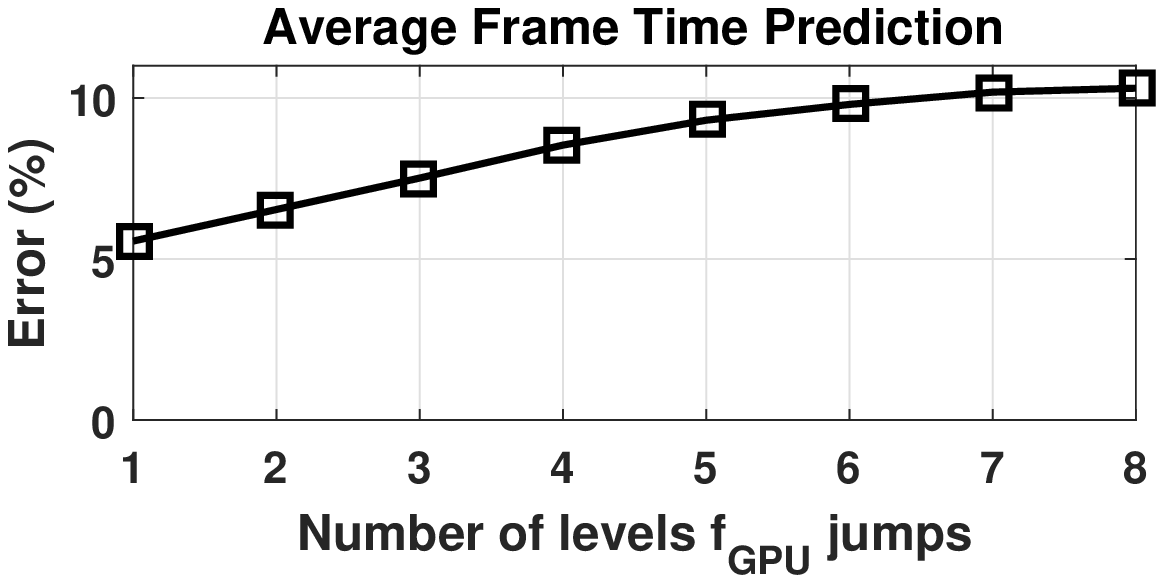}
	\caption{\textcolor{black}{Frame time prediction error in RenderingTest application for multiple frequency jumps.}}
	\label{fig:delta_F_up_down_rendering_test_Freq_jumps}
\end{figure}

To assess the accuracy of our sensitivity prediction,
we predict the change in frame time as a result of increasing (or decreasing) the frequency.
Then, we compute the frame time sensitivity using Equation~\ref{eq:frame_time_sensitivity}.
We start with changing the frequency by one level according the supported GPU frequencies listed in Figure~\ref{fig:data_set},
\emph{e.g.,} changing $f_{GPU}$ from $f_k=400$ MHz to
$f_{new}=444$ MHz or $f_{new}=355$ MHz.
Figure~\ref{fig:RLS_all_ave_trace_RenderingEngine_comp_actual_tF_up_to_pred_tF_up} shows
the predicted and actual frame time when the new frequency $f_\mathrm{new}$ is one level higher.
\textcolor{black}{The mean absolute percentage error for this prediction is $5.4\%$.
We observe the same result when $f_\mathrm{new}$ is one level lower. }
One might argue that the high prediction accuracy is only due to single frequency jumps like $400$ MHz to $444$ MHz.
Therefore, we also repeat our experiments for multiple frequency jumps.
For example, if current frequency is $200$ MHz, then a frequency jump of three
implies $f_\mathrm{new}$ is $311$ MHz.
\textcolor{black}{Figure~\ref{fig:delta_F_up_down_rendering_test_Freq_jumps} shows that the accuracy indeed degrades, but even when the number of frequency levels is eight (maximum allowed on Minnowboard), the error is less than $10\%$.
In practice, the frequency level changes in DTPM algorithms is not performed drastically from lowest to highest, but in smaller steps
leading to higher accuracy.}

\begin{figure}[t]
    \centering
    \includegraphics[width=\linewidth]{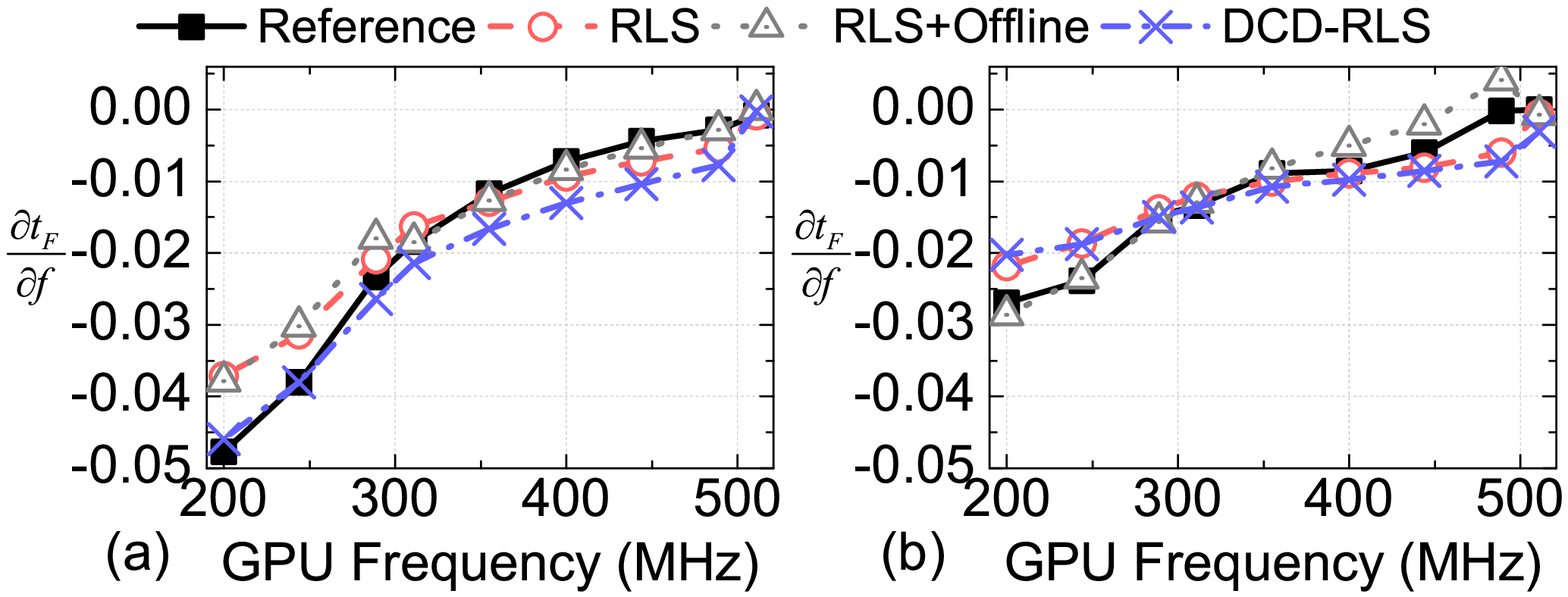}
    \caption{\textcolor{black}{Sensitivity of frame time with respect to frequency for (a) RenderingTest and (b) Art3 applications.}}
    \label{fig:frametime_derivative}
    \vspace{-2mm}
\end{figure}

\noindent \textbf{Accuracy of the Partial Derivative of Frame Time with Respect to Frequency:}
\textcolor{black}{We present the accuracy in predicting the derivative of frame time with respect to GPU frequency
for the RenderingTest and Art3 applications in Figure~\ref{fig:frametime_derivative}(a) and (b), respectively.
Each plot shows the derivative values for the reference, RLS, RLS+Offline, and DCD-RLS techniques.
We compute the derivative using Lagrange's polynomial method with change in frequency one level higher and one level lower,
as given by Equation~\ref{eq:lag_derivative}.
As seen from Figure~\ref{fig:frametime_derivative}(a), the slope starts with a negative value and then diminishes to zero on increasing frequency.
This is consistent with the observation in Figure~\ref{fig:counter_comp_all_freq}(a).
The normalized root mean squared error in the derivative prediction for RenderingTest application
using RLS, RLS+Offline, and DCD-RLS are $6.8\%$, $6.9\%$, and $5.9\%$, respectively.
These results indicate high accuracy for the derivative prediction,
with the RLS and DCD-RLS having an additional advantage of
performing the prediction completely online without using frequency dependent counters. This eliminates an extra step of predicting the derivative of the counter with respect to frequency.}
\textcolor{black}{In addition to running the RenderingTest application, we ran Art3 as well to measure frame time sensitivity.
Figure~\ref{fig:frametime_derivative}(b) shows that
the predicted derivative of frame time with respect to GPU frequency follows the reference values closely.
In particular, the normalized root mean squared error in the derivative prediction for Art3 application
using RLS, RLS+Offline, and DCD-RLS algorithms are $6.6\%$, $4.9\%$, and $8\%$, respectively.
Off note, the derivative values for Art3 application are smaller than the RenderingTest application, because Art3
is a memory bound graphics application.}

%% file: tferror_high_explaination.tex
We observe that the games BrainItOn and Jetski require extensive user interaction, which leads to fast changes in the frame time.
This makes the tracking of the rapidly changing frame time difficult and results in a mean error of $12\%$ and $10\%$, respectively.
Nonetheless, both these applications have low median absolute errors of $6.5\%$ and $1.3\%$,
which suggests that the error is not high for majority of time intervals.

%% file: multi_apps.tex
\noindent 
\textbf{Frame Time Prediction for Concurrent Application:}
Newer generation of mobile platforms using Android 7.1 have added support for running multiple applications using split-screen.
Therefore, it is important to also validate the performance model on these newer generation of platforms and multiple application scenarios.
For this experiment, we employ the Moto-X pure edition smartphone running Android 7.1 on a Qualcomm Snapdragon 808 SoC. 
We split the screen into two parts as top and bottom.
Then, we run a YouTube application in the bottom part of the screen and play the Chain reaction game simultaneously on the top part of the screen.
Figure~\ref{fig:MotoX_multiApps} shows the reference and predicted frame times for this multiple application scenario.
The proposed RLS algorithm achieves 8\% frame time prediction error.



%% file: time_method_motox.tex
There are many benefits of the online performance model compared to offline evaluation.
For example, in our case, the online modeling methodology reduced the characterization and model tuning effort
from several months to a few days for the Moto-X smartphone.
Similarly, the mobile platforms are expected to deliver good performance for any new applications that were created after the product launch.
Therefore, the online modeling technique enables adaptation to the new workloads without costly repetition of the workload characterization by the platform designers. Finally, our approach is easily portable 
and independent of any vendor and architecture. 

%% file: ar_lms.tex
%
In this section, we compare our approach to a tenth-order autoregressive (AR) model
which learns the model parameters using Normalized Least Mean Square (LMS) algorithm~\cite{dietrich2010lms}.
%
%
%
We first observe that LMS algorithm is slower to converge than RLS.
For example, Figure~\ref{fig:LMS_comp} shows that our approach converges to optimal model coefficients
in 50$ms$ while running the Icestorm application.
In contrast, the LMS approach takes 1.6$s$ to converge while running the same workload.
In general, the optimal model coefficient targets also change at runtime as the application phases change dynamically.
The convergence of the LMS approach is slow due to the tenth-order AR model, which takes the first ten samples to do the initial learning.
However, the convergence time of our approach varies between 50$ms$ to 0.3$s$, while LMS takes in the order of seconds.
We also note that the AR model can predict the frame time, but it
cannot predict the partial derivative of GPU performance with respect to frequency,
since it does not have a frequency term.
%
%
Therefore, our approach can directly provide the frequency sensitivity data to dynamic power management algorithms unlike the existing AR model~\cite{dietrich2010lms}.
Furthermore, fast convergence enables quick response to the dynamic changes in the workload.
%
%

%% file: dpm.tex
Our performance model can be used with a large variety of power management
algorithms that can optimize for system objectives, such as performance under a power budget~\cite{Zhang2016Maximizing, Gupta2017Dynamic} and energy~\cite{Gupta2017Dypo}.
In this section, we demonstrate the application of the proposed GPU performance model for
minimizing the energy consumption subject to a minimum frame rate constraint of 60 FPS.
At each control interval, we use the proposed GPU performance model to predict the frame time at all the frequencies supported in the platform.
Then, we select the frequency that leads to the smallest energy consumption, while meeting the minimum frame time constraint for the next interval.
To evaluate the effectiveness of our approach, we compare our results to an Oracle-based policy that precisely knows what the frequency in the next interval should be.
We obtain this information by running each frame at each supported frequency before this experiment.
Obviously, Oracle-based policy is not practical, but it provides the optimal results as a comparison point.
In addition to Oracle, we also compare our approach against the Linux Ondemand governor, which is used in many commercial products~\cite{pallipadi2006ondemand}.

For this experiment, we run industrial gaming workloads and our custom applications\footnote{
These workloads include games, such as, Fruit Ninja, Angry Birds, Jungle Run, Angry Bots, and Shark Dash, running on Intel core i5 6$^{th}$ generation platform.
We refer to these games as Workload 1-8 in the plot for confidentiality following the request from Intel.}. 
\input{tf_error_config2data.tex}
Figure~\ref{fig:DPM_bar} shows the energy consumption achieved by the Ondemand governor and the proposed RLS-based algorithm. 
The optimal energy value achieved by Oracle is shown by the dotted red line, 
and the other results are normalized to that of the Oracle-based policy. 

Workloads-6 to 8 and our custom applications have light-load graphics processing requirements. 
Consequently, these applications have low GPU utilization and can achieve the 60 FPS frame rate target with small GPU frequencies.
Our algorithm successfully chooses the right GPU frequency and matches the Oracle-based policy, as expected.
The Ondemand governor, which makes its decisions based on the GPU utilization, chooses small frequencies. 
As a result, it can also achieve the minimum energy consumption. 

Unlike the light-load graphics applications, the frame rate target cannot be achieved with 
lower GPU frequencies while running Workloads-1 to 5. 
These workloads are heavy to medium-load graphics games that result in high GPU utilization. 
In this case, high GPU utilization makes the Ondemand governor choose large frequencies. 
As a result, its energy consumption is 1.3$\times$-2.6$\times$ larger than the minimum energy achieved by the Oracle-based policy.
In contrast, our RLS-based approach can successfully choose the optimal operating frequencies due to its high accuracy.
Consequently, the energy consumption of our approach is within 1.06$\times$ of the optimal value.

Overall, our RLS-based policy leads to only 3\% higher average energy consumption compared to the Oracle-based policy.
In contrast, the Ondemand governor has 1.3$\times$-2.6$\times$ larger energy consumption under heavy workloads.
On average, our RLS-based policy provides about 43\% lower energy consumption compared to the Ondemand governor while achieving the same frame rate.

%% file: tf_error_config2data.tex
Out of these interactive games, the first five workloads have frame time error less than 4\%, while
the remaining workloads 6-8 have higher frame time errors of more than 10\%. 

%% file: overhead.tex
\subsection{Overhead Analysis} \label{sec:overhead_analysis}
\begin{figure}[h]
	\centering	\includegraphics[width=0.9\linewidth]{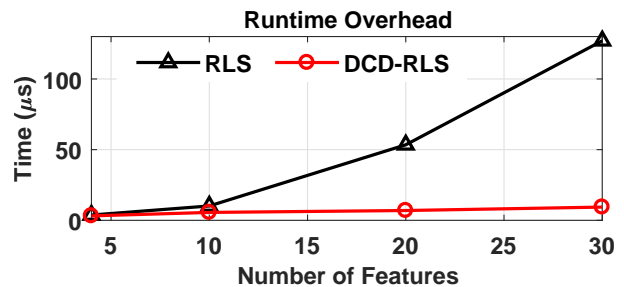}
	\caption{\textcolor{black}{Overhead time as a function of the number of features for the RLS and DCD-RLS algorithm.}}
	\label{fig:overhead_comparison}
\end{figure}

\textcolor{black}{We measure the overhead of the proposed approach by
instrumenting the start and end times of each of the RLS iterations,
including the feature data preparation step.
Then, we measure the time for the proposed frame time prediction mechanism
running on the Minnowboard platform.
Figure~\ref{fig:overhead_comparison} demonstrates the difference in the runtime overheads of the RLS and DCD-RLS algorithms in each iteration.
When the number of features are four, the overhead time of the RLS and DCD-RLS algorithms are 3.8$\mu$s and 3.2$\mu$s, respectively.
As the number of features increase to 20, the runtime overhead of the RLS algorithm becomes much larger than DCD-RLS.
More precisely, for 20 features, the RLS algorithm has the runtime overhead of
53.4$\mu$s, while the DCD-RLS algorithm has 7.6$\times$ smaller overhead of 7$\mu$s.
This experiment demonstrates that the proposed RLS technique has very low overhead for a small number of features.
When the number of features are large and lowering the overhead time is critical, DCD-RLS is a viable alternative to the proposed RLS algorithm.}

%% file: conclusion_future_research.tex
\section{Conclusion and Future Work} \label{sec:conclusion}

\textcolor{black}{In this paper, we propose an online performance modeling methodology for graphics cores.
The proposed methodology combines offline data collection and online learning using RLS algorithm.
Online learning of the model coefficients enables adapting to unknown workloads by
eliminating the need for costly offline training.}
\textcolor{black}{Extensive evaluations on an experimental platform using common GPU benchmarks resulted in
average mean absolute errors of $4.2\%$ in frame time and $6.7$\% in frame time sensitivity prediction.
Furthermore, we experimentally showed that the proposed high accuracy performance model
could be successfully employed by an dynamic power management algorithm
that minimizes energy consumption under a performance constraint.}

\noindent \textbf{Acknowledgments:} \textcolor{black}{This work was supported partially by Semiconductor Research Corporation (SRC) task 2721.001 and National Science Foundation (NSF) grant CNS-1526562.} 

%% file: appendix1.tex
\section*{\textcolor{black}{Appendix-1}} \label{sec:appendix1}

\noindent \textcolor{black}{\textbf{Derivation of Equation~\ref{eq:general_model_freq_indep}:}
In this section, we explain the steps to achieve the model structure that does not use the frequency dependent hardware counters.
We start with the general frame time model structure from Equation~\ref{eq:general_model} and
substitute Equation~\ref{eq:change_in_counters} in it.
}

\textcolor{black}{
\vspace{-2mm}
\small
\begin{equation}\label{eq:appendix1}
  \Delta t_{F,k}(f_k,\mathbf{x}_k(f_k)) \hspace{-0.02in} \approx \hspace{-0.02in} a_0 t_{F,k-1} \hspace{-0.04in} \left (\hspace{-0.02in} \frac{f_{k-1}}{f_{k}} - 1 \hspace{-0.02in} \right ) \hspace{-0.02in} + \hspace{-0.02in} \sum_{i=1}^N \hspace{-0.02in} a_i \hspace{-0.02in}
\left ( \frac{ \partial x_{i,k} }{\partial f}  \Delta f_k \hspace{-0.03in}+ \hspace{-0.03in} \frac{ \partial x_{i,k} }{\partial C} \Delta C \right )
\end{equation}
\normalsize
}

\noindent
\textcolor{black}{
Then, we separate the terms for the change in counters due to frequency $f$ and complexity $C$.
}
\textcolor{black}{
\vspace{-2mm}
\small
\begin{multline}\label{eq:appendix2}
  \Delta t_{F,k}(f_k,\mathbf{x}_k(f_k)) \hspace{-0.02in} = \hspace{-0.02in} a_0 t_{F,k-1} \hspace{-0.04in} \left (\hspace{-0.02in} \frac{f_{k-1}}{f_{k}} - 1 \hspace{-0.02in} \right ) +  \left ( \sum_{i=1}^N \hspace{-0.02in} a_i
 \frac{ \partial x_{i,k} }{\partial f}\right )  \Delta f_k \\ + \hspace{-0.02in} \sum_{i=1}^N \hspace{-0.02in} a_i \left (  \frac{ \partial x_{i,k} }{\partial C} \Delta C \right )
\end{multline}
\normalsize
}

\noindent
\textcolor{black}{ We combine the term $\sum_{i=1}^N a_i \frac{ \partial x_{i,k} }{\partial f}$ into a single
model coefficient $b_1$ that is learned at runtime.
}
\textcolor{black}{
\vspace{-2mm}
\small
\begin{equation}\label{eq:appendix3}
  \Delta t_{F,k}(f_k,\mathbf{x}_k(f_k)) \hspace{-0.02in} = \hspace{-0.02in} a_0 t_{F,k-1} \hspace{-0.04in} \left (\hspace{-0.02in} \frac{f_{k-1}}{f_{k}} - 1 \hspace{-0.02in} \right ) \hspace{-0.02in} + \hspace{-0.02in} b_1  \Delta f_k  + \hspace{-0.02in} \sum_{i=1}^{N} \hspace{-0.02in} a_i \left (  \frac{ \partial x_{i,k} }{\partial C} \Delta C \right )
\end{equation}
\normalsize
}

\noindent
\textcolor{black}{The change in the counters that are frequency independent can be written as $ \Delta x_k = \frac{ \partial x_{k} }{\partial C} \Delta C$. As a result, we can change the summation in the third term to only include the frequency independent counters without loss of
generality.
}

\textcolor{black}{
\vspace{-2mm}
\small
\begin{equation}\label{eq:appendix4}
  \Delta t_{F,k}(f_k,\mathbf{x}_k(f_k)) \hspace{-0.02in} = \hspace{-0.02in} a_0 t_{F,k-1} \hspace{-0.04in} \left (\hspace{-0.02in} \frac{f_{k-1}}{f_{k}} - 1 \hspace{-0.02in} \right ) \hspace{-0.02in} + \hspace{-0.02in} b_1  \Delta f_k  + \hspace{-0.02in} \sum_{i=1}^{N_\mathrm{indep}} \hspace{-0.02in} b_{i+1} \Delta x_{i,k}
\end{equation}
\normalsize
}

\noindent
\textcolor{black}{Finally, we perform a change in variables for the model coefficients to represent $b$ with $a$ and obtain
Equation~\ref{eq:general_model_freq_indep}.
}

\textcolor{black}{
\vspace{-2mm}
\small
\begin{equation}\label{eq:appendix4}
\Delta t_{F,k}(f_k,\mathbf{x}_k(f_k)) \hspace{-0.02in} = \hspace{-0.02in} a_0 t_{F,k-1} \hspace{-0.04in} \left (\hspace{-0.02in} \frac{f_{k-1}}{f_{k}} - 1 \hspace{-0.02in} \right ) \hspace{-0.02in} + \hspace{-0.02in} a_1  \Delta f_k  + \hspace{-0.02in} \sum_{i=1}^{N_\mathrm{indep}} \hspace{-0.02in} a_{i+1} \Delta x_{i,k} \nonumber
\end{equation}
\normalsize
}

%% file: main.bbl
\begin{thebibliography}{43}
\providecommand{\natexlab}[1]{#1}
\providecommand{\url}[1]{#1}
\csname url@samestyle\endcsname
\providecommand{\newblock}{\relax}
\providecommand{\bibinfo}[2]{#2}
\providecommand{\BIBentrySTDinterwordspacing}{\spaceskip=0pt\relax}
\providecommand{\BIBentryALTinterwordstretchfactor}{4}
\providecommand{\BIBentryALTinterwordspacing}{\spaceskip=\fontdimen2\font plus
\BIBentryALTinterwordstretchfactor\fontdimen3\font minus
  \fontdimen4\font\relax}
\providecommand{\BIBforeignlanguage}[2]{{%
\expandafter\ifx\csname l@#1\endcsname\relax
\typeout{** WARNING: IEEEtranSN.bst: No hyphenation pattern has been}%
\typeout{** loaded for the language `#1'. Using the pattern for}%
\typeout{** the default language instead.}%
\else
\language=\csname l@#1\endcsname
\fi
#2}}
\providecommand{\BIBdecl}{\relax}
\BIBdecl

\bibitem[AnandTech()]{AnandTech_ARM_MALI}
AnandTech. {ARM's Mali Midgard Architecture Explored}.
  \url{https://www.anandtech.com/show/8234/arms-mali-midgard-architecture-explored/4}.

\bibitem[{App Tornado}()]{AppTornadoApp}
{App Tornado}. {App Brain}. ~\url{http://www.appbrain.com/}, accessed July 20,
  2016.

\bibitem[Ayoub et~al.(2011)]{ayoub2011OS}
R.~Z. Ayoub \emph{et~al.}, ``{OS-level Power Minimization under Tight
  Performance Constraints in General Purpose Systems},'' in \emph{Proc. of the
  Intl. Symp. on Low-power Electronics and Design}, 2011, pp. 321--326.

\bibitem[Benini et~al.(2000)Benini, Bogliolo, and De~Micheli]{Benini2000Survey}
L.~Benini, A.~Bogliolo, and G.~De~Micheli, ``{A Survey of Design Techniques For
  System-Level Dynamic Power Management},'' \emph{IEEE Trans. Very Large Scale
  Integr. (VLSI) Syst.}, vol.~8, no.~3, pp. 299--316, 2000.

\bibitem[Dietrich and Chakraborty(2014)]{dietrich2014lightweight}
B.~Dietrich and S.~Chakraborty, ``{Lightweight Graphics Instrumentation for
  Game State-Specific Power Management in Android},'' \emph{Multimedia
  Systems}, vol.~20, no.~5, pp. 563--578, 2014.

\bibitem[Dietrich et~al.(2010)]{dietrich2010lms}
B.~Dietrich \emph{et~al.}, ``{LMS-based Low-complexity Game Workload Prediction
  for DVFS},'' in \emph{Proc. of the Intl. Conf. on Comp. Design}, 2010, pp.
  417--424.

\bibitem[Faith(1999)]{faith1999direct}
R.~Faith, ``{The Direct Rendering Manager: Kernel Support for the Direct
  Rendering Infrastructure},'' 1999.

\bibitem[Farrar and Glauber(1967)]{farrar1967multicollinearity}
D.~E. Farrar and R.~R. Glauber, ``{Multicollinearity in Regression Analysis:
  the Problem Revisited},'' \emph{The Review of Economic and Statistics,
  JSTOR}, pp. 92--107, 1967.

\bibitem[Friedman et~al.(2001)Friedman, Hastie, and
  Tibshirani]{friedman2001elements}
J.~Friedman, T.~Hastie, and R.~Tibshirani, \emph{{The Elements of Statistical
  Learning}}.\hskip 1em plus 0.5em minus 0.4em\relax {Springer Series in
  Statistics, Berlin}, 2001, vol.~1.

\bibitem[Gu et~al.(2006)Gu, Chakraborty, and Ooi]{gu2006games}
Y.~Gu, S.~Chakraborty, and W.~T. Ooi, ``{Games are up for DVFS},'' in
  \emph{Proc. of the Design Automation Conf.}, 2006, pp. 598--603.

\bibitem[Gupta et~al.(2016{\natexlab{a}})Gupta, Campbell, Ogras, Ayoub,
  Kishinevsky, Paterna, and Gumussoy]{Gupta2016Adaptive}
U.~Gupta, J.~Campbell, U.~Y. Ogras, R.~Ayoub, M.~Kishinevsky, F.~Paterna, and
  S.~Gumussoy, ``{Adaptive Performance Prediction for Integrated GPUs},'' in
  \emph{Proc. of the Intl. Conf. on Computer-Aided Design}, 2016, p.~61.

\bibitem[Gupta et~al.(2016{\natexlab{b}})Gupta, Korrapati, Matturu, and
  Ogras]{Gupta2016Generic}
U.~Gupta, S.~Korrapati, N.~Matturu, and U.~Y. Ogras, ``{A Generic Energy
  Optimization Framework for Heterogeneous Platforms Using Scaling Models},''
  \emph{Microprocessors and Microsystems}, vol.~40, pp. 74--87, 2016.

\bibitem[Gupta et~al.(2017{\natexlab{b}})Gupta, Patil, Bhat, Mishra, and
  Ogras]{Gupta2017Dypo}
U.~Gupta, C.~A. Patil, G.~Bhat, P.~Mishra, and U.~Y. Ogras, ``{DyPO: Dynamic
  Pareto Optimal Configuration Selection for Heterogeneous MpSoCs},'' \emph{ACM
  Tran. on Embedded Comp. Sys. (to appear)}, Oct 2017.

\bibitem[Gupta et~al.(2017{\natexlab{a}})]{Gupta2017Dynamic}
U.~Gupta \emph{et~al.}, ``{Dynamic Power Budgeting for Mobile Systems Running
  Graphics Workloads},'' \emph{IEEE Trans. on Multi-Scale Comp. Sys.}, 2017.

\bibitem[Hoerl and Kennard(1970)]{hoerl1970ridge}
A.~E. Hoerl and R.~W. Kennard, ``{Ridge regression: Biased Estimation for
  Nonorthogonal Problems},'' \emph{Technometrics, Taylor \& Francis Group},
  vol.~12, no.~1, pp. 55--67, 1970.

\bibitem[{Intel Corp.}(2015)]{Intel2015Open}
{Intel Corp.}, \emph{{Open Source HD Graphics Programmers' Reference Manual}},
  June 2015.

\bibitem[{Intel Corp.}({\natexlab{b}})]{IntelIntelb}
------. {Intel GPU Tools}. {http://01.org/linuxgraphics/gfx-docs/igt/},
  accessed July 20, 2016.

\bibitem[{Intel Corp.}({\natexlab{a}})]{IntelMinnowboard}
------. {Minnowboard}. {http://www.minnowboard.org/}, accessed July 20, 2016.

\bibitem[Ismail and Principe(1996)]{ismail1996equivalence}
M.~Ismail and J.~Principe, ``{Equivalence Between RLS Algorithms and the Ridge
  Regression Technique},'' in \emph{Proc. of the Conf. on Signals, Sys. and
  Comp.}, 1996, pp. 1083--1087.

\bibitem[Jin et~al.(2015)Jin, He, and Liu]{jin2015towards}
T.~Jin, S.~He, and Y.~Liu, ``{Towards Accurate GPU Power Modeling for
  Smartphones},'' in \emph{Proc. of the Workshop on Mobile Gaming}, 2015, pp.
  7--11.

\bibitem[Kadjo et~al.(2015)Kadjo, Ayoub, Kishinevsky, and
  Gratz]{kadjo2015control}
D.~Kadjo, R.~Ayoub, M.~Kishinevsky, and P.~V. Gratz, ``{A Control-Theoretic
  Approach for Energy Efficient CPU-GPU Subsystem in Mobile Platforms},'' in
  \emph{Proc. of the Design Autom. Conf.}, 2015, pp. 62:1--62:6.

\bibitem[Ma et~al.(2014)Ma, Wang, and Wang]{ma2014dppc}
K.~Ma, X.~Wang, and Y.~Wang, ``{DPPC: Dynamic Power Partitioning and Control
  for Improved Chip Multiprocessor Performance},'' \emph{IEEE Trans. on Comp.},
  vol.~63, no.~7, pp. 1736--1750, 2014.

\bibitem[{Mathworks}()]{MathworksMATLABb}
{Mathworks}. {MATLAB System Identification Toolbox}.
  {https://www.mathworks.com/products/sysid.html}.

\bibitem[Mendel(1995)]{Mendel1995Lessons}
J.~M. Mendel, \emph{{Lessons in Estimation Theory for Signal Processing,
  Communications, and Control}}.\hskip 1em plus 0.5em minus 0.4em\relax Pearson
  Educ., 1995.

\bibitem[{Mishra, Nikita and Zhang, Huazhe and Lafferty, John D and Hoffmann,
  Henry}(2015)]{mishra2015probabilistic}
{Mishra, Nikita and Zhang, Huazhe and Lafferty, John D and Hoffmann, Henry},
  ``{A Probabilistic Graphical Model-Based Approach for Minimizing Energy Under
  Performance Constraints},'' in \emph{ACM SIGARCH Computer Architecture News},
  vol.~43, no.~1, 2015, pp. 267--281.

\bibitem[Mochocki et~al.(2006)Mochocki, Lahiri, and Cadambi]{mochocki2006power}
B.~Mochocki, K.~Lahiri, and S.~Cadambi, ``{Power Analysis of Mobile 3D
  Graphics},'' in \emph{Proc. of the Design, Autom. and Test in Europe Conf.},
  2006, pp. 502--507.

\bibitem[Motorola()]{motoX}
Motorola. {Moto X Pure Edition Smartphone}.
  \url{https://www.motorola.com/us/products/moto-x-pure-edition}.

\bibitem[Nagasaka et~al.(2010)]{nagasaka2010statistical}
H.~Nagasaka \emph{et~al.}, ``{Statistical Power Modeling of GPU Kernels using
  Performance Counters},'' in \emph{Proc. of the Intl. Green Computing Conf.},
  2010, pp. 115--122.

\bibitem[{National Instr.}()]{NationalNI}
{National Instr.} {NI USB-6289}.
  \url{http://sine.ni.com/nips/cds/view/p/lang/en/nid/209154}.

\bibitem[Ogras et~al.(2013)Ogras, Ayoub, Kishinevsky, and
  Kadjo]{Ogras2013Managing}
U.~Y. Ogras, R.~Z. Ayoub, M.~Kishinevsky, and D.~Kadjo, ``{Managing Mobile
  Platform Power},'' in \emph{Proc. of Intl. Conf. on Computer-Aided Design},
  2013, pp. 161--162.

\bibitem[Pallipadi and Starikovskiy(2006)]{pallipadi2006ondemand}
V.~Pallipadi and A.~Starikovskiy, ``{The Ondemand Governor},'' in \emph{Proc.
  of the Linux Symp.}, vol.~2, 2006.

\bibitem[Paterna et~al.(2017)Paterna, Gupta, Ayoub, Ogras, and
  Kishinevsky]{paterna2017adaptive}
F.~Paterna, U.~Gupta, R.~Ayoub, U.~Y. Ogras, and M.~Kishinevsky, ``{Adaptive
  Performance Sensitivity Model to Support GPU Power Management},'' in
  \emph{Proc. of the Workshop on Autotuning and Adaptivity Approaches for
  Energy Efficient HPC Sys.}, 2017, p.~5.

\bibitem[Pathania et~al.(2015)Pathania, Irimiea, Prakash, and
  Mitra]{Pathania2015Power}
A.~Pathania, A.~E. Irimiea, A.~Prakash, and T.~Mitra, ``{Power-Performance
  Modelling of Mobile Gaming Workloads on Heterogeneous MPSoCs},'' in
  \emph{Proc. of the Design Autom. Conf.}, 2015, pp. 201:1--201:6.

\bibitem[Pothukuchi et~al.(2016)Pothukuchi, Ansari, Voulgaris, and
  Torrellas]{pothukuchi2016using}
R.~P. Pothukuchi, A.~Ansari, P.~Voulgaris, and J.~Torrellas, ``{Using Multiple
  Input, Multiple Output Formal Control to Maximize Resource Efficiency in
  Architectures},'' in \emph{Proc. of the Intl. Symp. on Computer Architecture
  (ISCA)}, 2016, pp. 658--670.

\bibitem[Sayed(2003)]{sayed2003fundamentals}
A.~H. Sayed, \emph{{Fundamentals of Adaptive Filtering}}.\hskip 1em plus 0.5em
  minus 0.4em\relax John Wiley \& Sons, 2003.

\bibitem[Sayed(2011)]{sayed2011adaptive}
------, \emph{{Adaptive Filters}}.\hskip 1em plus 0.5em minus 0.4em\relax John
  Wiley \& Sons, 2011.

\bibitem[Singh and Bhadauria(2009)]{singh2009finite}
A.~K. Singh and B.~Bhadauria, ``{Finite Difference Formulae for Unequal
  Sub-intervals using Lagrange’s Interpolation Formula},'' \emph{Int. J.
  Math. Anal}, vol.~3, no.~17, p. 815, 2009.

\bibitem[Strang(2007)]{strang2007computational}
G.~Strang, \emph{{Computational Science and Engineering}}.\hskip 1em plus 0.5em
  minus 0.4em\relax Wellesley-Cambridge Press, 2007, vol. 791.

\bibitem[Varatkar and Marculescu(2004)]{varatkar2004chip}
G.~V. Varatkar and R.~Marculescu, ``{On-chip Traffic Modeling and Synthesis for
  MPEG-2 Video Applications},'' \emph{IEEE Trans. on Very Large Scale Integr.
  (VLSI) Sys.}, vol.~12, no.~1, pp. 108--119, 2004.

\bibitem[Wang et~al.(2016)]{wang2016pareto}
X.~Wang \emph{et~al.}, ``{A Pareto-Optimal Runtime Power Budgeting Scheme for
  Many-Core Systems},'' \emph{Microprocessors and Microsystems}, vol.~46, pp.
  136--148, 2016.

\bibitem[Wang et~al.(2011)Wang, Ma, and Wang]{wang2011adaptive}
X.~Wang, K.~Ma, and Y.~Wang, ``{Adaptive Power Control With Online Model
  Estimation for Chip Multiprocessors},'' \emph{IEEE Trans. on Parallel and
  Distributed Sys.}, vol.~22, no.~10, pp. 1681--1696, 2011.

\bibitem[Zakharov et~al.(2008)Zakharov, White, and Liu]{Zakharov2008Low}
Y.~V. Zakharov, G.~P. White, and J.~Liu, ``{Low-complexity RLS Algorithms Using
  Dichotomous Coordinate Descent Iterations},'' \emph{IEEE Trans. on Signal
  Proc.}, vol.~56, no.~7, pp. 3150--3161, 2008.

\bibitem[Zhang and Hoffmann(2016)]{Zhang2016Maximizing}
H.~Zhang and H.~Hoffmann, ``{Maximizing Performance Under a Power Cap: A
  Comparison of Hardware, Software, and Hybrid Techniques},'' in \emph{Proc. of
  the Intl. Conf. on Arch. Support for Programming Languages and Operating
  Sys.}, 2016, pp. 545--559.

\end{thebibliography}
